\documentclass[11pt,a4paper]{article}
\usepackage{jheppub}
\usepackage{etoolbox}
\makeatletter
\patchcmd{\maketitle}{\@fpheader}{}{}{}
\makeatother
\usepackage{graphicx}
\usepackage{amssymb}
\usepackage{epstopdf}
\usepackage{booktabs}
\usepackage{subfigure}
\usepackage{rotating}
\usepackage{lineno}
\title{The dependency of boosted tagging algorithms on the event colour structure}
\author[a]{K. Joshi} 
\author[a,b]{A.D. Pilkington}
\author[b]{M. Spannowsky}
\affiliation[a]{School of Physics \& Astronomy, University of Manchester, Manchester M13 9PL, UK.}
\affiliation[b]{Institute of Particle Physics Phenomenology,
  University of Durham, Durham DH1 3LE, UK.}
\abstract{
The impact of event colour structure on the performance of the \jh{}, \cms{},  \heptt{} and \nsub{} algorithms is investigated by studying colour singlet and colour octet resonances decaying to top-quark pairs. Large differences in top-tagging efficiency are observed due to the different colour charge of each resonance. These differences are quantified as a function of the algorithm parameters, the jet size parameter and the probability to misidentify light quarks and gluons as top candidates. We suggest that future experimental searches would benefit from optimising the choice of algorithm parameters in order to minimise this source of model dependency.
}

\newcommand{\heptt}{\textsc{heptoptagger}}{}
\newcommand{\jh}{\textsc{johns-hopkins}}{}
\newcommand{\cms}{\textsc{cms}}{}
\newcommand{\nsub}{\textsc{n-subjettiness}}{}

\begin{document}
\maketitle



\section{Introduction}
\label{sec:case}

The benefit of using boosted object reconstruction in searches for heavy resonances has only been fully appreciated in recent years \cite{mike,butterworth,bdrs,boost2010,subreview,boost2011}. At the Large Hadron Collider (LHC),  electroweak-scale resonances can be produced with transverse momentum, $p_{\rm T}^{}$, much larger than the particle's mass, $m$, due to the large centre-of-mass energy of the colliding protons. The decay products of an object produced with $p_{\rm T}^{} \gg m$ tend to be collimated in the lab frame and confined to a small area of the detector. This reduces combinatorial problems in reconstructing the object in question, by collecting all radiation from the decay products into one `fat' jet. Jet substructure techniques, which study the energy distribution of the constituents inside the fat jet, are then necessary to distinguish between those jets originating from boosted objects and the QCD-induced backgrounds from light quark and gluon jets.

The reconstruction of highly boosted top quarks is a particularly well documented physics case in which jet substructure techniques are indispensable if a New Physics signal is to be extracted from the large QCD background \cite{had_resonances,semi_resonances}. 
Several boosted top quark reconstruction algorithms have been proposed in the literature (for an overview see \cite{topreview}). These so-called `top-taggers' make use of two classes of observable. The first class are well established event shape observables that have been elevated to jet shape observables \cite{thaler_wang,leandro1,nsub,treeless}.
The most prominent example of a jet shape is the jet mass, upon which all top-tagging algorithms rely to some extent in order to discriminate the top signal from light quark and gluon backgrounds. 
The second class of observable is the clustering history of the jet constituents. 
Infrared-safe sequential jet algorithms introduce a metric that determines the 
ordering for jet constituents to be combined into subjets and, eventually, into the fat jet. 
As the radiation profile of quarks and gluons is different in comparison to the profile produced by the decay of a boosted top quark, tracing backward through the recombination history and comparing the energy sharing between the subjets allows us to identify signal and background jets \cite{hopkins,cmstagger,pruning1,heptop}. 

These top-tagging algorithms, and boosted algorithms in general, only use information related to the fat-jet constituents and implicitly assume that the overall hadronic event activity is of no major importance. However, radiation from the initial-state, the hard interaction and the underlying event can enter the fat-jet. The efficiency for tagging boosted objects will therefore depend upon the underlying colour structure in the event and, in particular, the colour of the decaying resonance, a fact that has been largely ignored in the literature thus far. 

In this article, we study the effect of resonance colour on the efficiency of well established top-tagging algorithms. 
We focus on $s$-channel resonances with different colour charge, i.e. a colour-singlet Kaluza-Klein (KK) photon, $\gamma_{KK}^{}$, and a colour-octet KK gluon, $G_{KK}^{}$. The two resonances produce very different radiation patterns. The colour octet can itself radiate gluons and the top quarks are colour connected with the initial state, whereas the initial and final states are not inter-connected for a colour singlet resonance. The results we will present have direct relevance for ongoing searches for heavy colour singlet and colour octet resonances at the LHC \cite{cmssearch,atlasresonance}. Finally, although we focus on resonances that decay into $t\bar{t}$ pairs, the results we present should be qualitatively similar for boosted decays into other electroweak-scale resonances. 

The article is arranged as follows: 
In Sec. \ref{sec:toptag} we introduce the jet reconstruction and top-tagging algorithms. In Sec~\ref{sec:evgen}, we present the Monte Carlo event generator used to study top-tagging efficiencies. In Sec. \ref{results1} present the performance of the top-tagging algorithms for the different signal models. In Sec.~\ref{results2}, we discuss the experimental difficulties in establishing top-tagging performance in light of these new results. 


\section{Boosted top reconstruction}
\label{sec:toptag}


The \jh{} top-tagger \cite{hopkins} is a descendant of the two-body approach of \cite{bdrs}. In the first step, the energy sharing of subjet splittings is evaluated whilst browsing backwards through the recombination history. If a jet splitting ($j \to j_1 j_2$) produces two subjets that satisfy $\min (p_{\rm T,1}^{},p_{\rm T,2}^{}) / p_{\rm T,j}^{} > \delta_p$ (mass-drop) and $\Delta R_{j_1,j_2}>\delta_r$, the subjets are retained as candidates for a top-quark decay. The default parameters are chosen to be $\delta_p=0.12$ and $\delta_r=0.16$. If a mass drop is observed, each subjet is immediately tested for another mass drop. The fat-jet is considered as a top candidate if three or four subjets are kept after the double-mass-drop procedure. Kinematic cuts are then applied to these subjets; the summed subjet mass should be in a top-mass window, a subjet pair should reconstruct the $W$-boson mass within a given window and the helicity angle\footnote{The helicity angle is defined as the angle, measured in the rest frame of the reconstructed $W$, between the direction of the reconstructed top and one of the $W$ decay products.} for this pair should be below a given threshold ($\cos \theta_{h} < 0.7$).

The \cms{} top-tagger is a modification to the \jh{} algorithm \cite{cmstagger}. In the first step, the angular separation of the subjets must pass a $p_{\rm T}$-dependent criterion, i.e. $\Delta R_{j_1,j_2} > \delta_r - A \cdot p_{\rm T, j}$, with default values of $\delta_r=0.4$ and $A=0.0004$. At least three subjets have to be present after passing the double-mass-drop procedure and, among the three hardest subjets, the two subjets with the smallest invariant mass are required to be consistent with the $W$-boson mass. The top candidate is defined as the three or four hardest subjets.

The \heptt{} was proposed for the reconstruction of mildly boosted top quarks in a busy environment \cite{heptop}. The first step aims to find the complete hard substructure of the fat jet, irrespective of the expected number of top decay products, and does not stop after a certain number of mass-drop conditions are met. It browses through all branches of the jet recombination history comparing at every step the relative mass sharing of the $j_1j_2 \to j$ mergings. The condition $\min(j_1,j_2) < 0.8\cdot m_{j}$ determines if $j_1$ and $j_2$ are kept as candidates for a heavy object decay. 
Subjets with $m_j < 30$ GeV are not unclustered further and are stored as potential top decay products. In the second step, all three-subjet combinations are filtered, similar to \cite{bdrs}, and the combination that is closest to the top mass within a specific top mass window is considered to be the top candidate. The subjets of the top candidate are required to pass three criteria:
\begin{alignat}{5}
&0.2 <\arctan \frac{m_{13}}{m_{12}} < 1.3
\qquad \text{and} \quad
R_{\min}< \frac{m_{23}}{m_{123}} < R_{\max}
\notag \\
&R_{\min}^2 \left(1+\left(\frac{m_{13}}{m_{12}}\right)^2 \right) 
< 1-\left(\frac{m_{23}}{m_{123}} \right)^2
< R_{\max}^2 \left(1+\left(\frac{m_{13}}{m_{12}}\right)^2 \right)       
\quad \text{and} \quad 
\frac{m_{23}}{m_{123}} > R_\text{soft}
\notag \\
&R_{\min}^2\left(1+\left(\frac{m_{12}}{m_{13}}\right)^2 \right) 
< 1-\left(\frac{m_{23}}{m_{123}} \right)^2
< R_{\max}^2\left(1+\left(\frac{m_{12}}{m_{13}}\right)^2 \right)        
\quad \text{and} \quad 
\frac{m_{23}}{m_{123}}> R_\text{soft}.
\label{eq:heptop}
\end{alignat} 
The default values of the dimensionless mass window bounds are set to $R_{\rm{min}}=0.85 \cdot m_{W}^{}/m_{t}^{}$ and $R_{\rm{max}}= 1.15 \cdot m_{W}^{}/ m_{t}^{}$.

The \nsub{} top-tagger \cite{nsub}  is an adaption of the event shape observable N-jettiness \cite{njet}. The observable aims to quantify the degree to which jet radiation is aligned along specific subjet axes. Smaller values of $\tau_N$ correspond to $N$ or fewer energy deposits inside the fat jet, while large values of $\tau_N$ indicate more than $N$ energy deposits. After using the exclusive $k_{\rm T}$ algorithm \cite{kt_algo} to determine the candidate subjet directions, $\tau_N$ is calculated by
\begin{equation}
\tau_N = \frac{\sum_k p_{\rm T,k}^{} \min(\Delta R_{1,k},\Delta R_{2,k},...,\Delta R_{N,k})}{\sum_k p_{\rm T,k}^{} R_0}.
\end{equation}
The summation is over all jet constituents, $p_{\rm T,k}^{}$ is the transverse momentum of each jet constituent, $\Delta R_{A,k}$ is the distance in $\eta\times\phi$ space between the subjet axis $A$ and each jet consituent, and $R_0$ is the jet radius. The ratio $\tau_N/\tau_{N-1}$  provides a good separation of top jets from light quark and gluon jets. The parameters used to define top candidates are $\tau_N/\tau_{N-1}$ and the fat-jet mass.

\section{Monte Carlo event simulation} 
\label{sec:evgen}

To study the effect of colour flow on the top-tagging performance, we simulate the production of heavy gluons and heavy photons in proton-proton collisions at $\sqrt{s}=14$~TeV. The production of heavy gluons decaying to top-quark pairs ($ q\bar{q} \rightarrow G_{KK} \rightarrow t\bar{t}$) is simulated using the {\sc pythia 8} event generator, with the specific implementation documented in \cite{Ask:2011zs}. The gluon resonance couplings are chosen to be $g^q_v = 0.2$, $g^b_v = 0$ and $g^t_v = 3.6$, and the resonance mass is chosen to be 2~TeV. This results in a resonance width of $\Gamma/M = 0.2$ and a cross section~$\times$~branching-ratio of 1.1~pb at $\sqrt{s}=14$~TeV. Heavy photon production is simulated by changing the colour factors, $\alpha_s \rightarrow e_q^2 \alpha_{em}$, where $e_q$ is the quark electric charge. The couplings of the heavy photon to light quarks and top quarks are then adjusted to reproduce heavy gluon production rate and resonance width, respectively. 

The events are generated with the CTEQ5L parton distribution functions \cite{cteqpdf}. Parton showering, hadronisation and underlying event from multiple parton interactions are generated with the default tune to {\sc pythia 8.130} (Tune 1) \cite{pythia}. For final states containing top-quarks, it is possible to simulate events in {\sc pythia 8} with two different parton shower options: `wimpy showers' allow radiation up to the factorisation scale of the process, whereas `power' showers allow radiation up to the kinematic limit. We found that the former option gave a much better description for the jet activity in Standard Model (SM) $t\bar{t}$ events when compared to the ATLAS data \cite{ATLAS:2012al}  and `wimpy' showers are therefore used throughout this study. The parton shower comparison to the data is documented in Figure \ref{fig:atlaspower} of Appendix A.

The probability that an algorithm wrongly identifies jets originating from light quarks and gluons as top-candidates is also assessed, to allow comparisons between different top-tagging algorithms for a fixed mis-tag probability. This light quark and gluon background is also simulated using {\sc pythia 8} with the default underlying event tune.

\section{Performance of top-tagging algorithms}
\label{results1}

\subsection{Basic kinematic features of the signal events}

\begin{figure}[t]%
\centering
\mbox{
\subfigure[]{\includegraphics[width=0.5\textwidth]{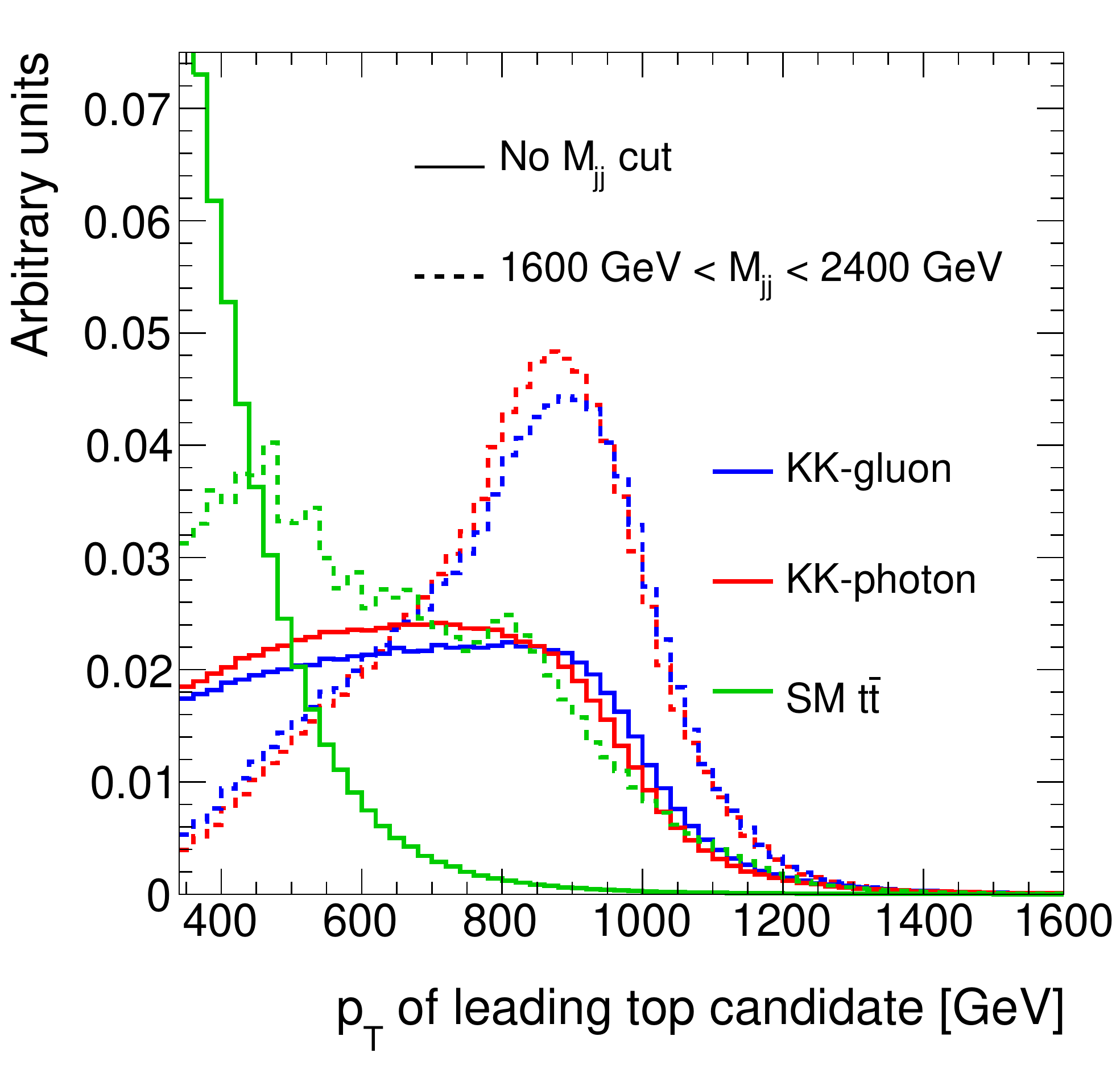} }%
\subfigure[]{\includegraphics[width=0.5\textwidth]{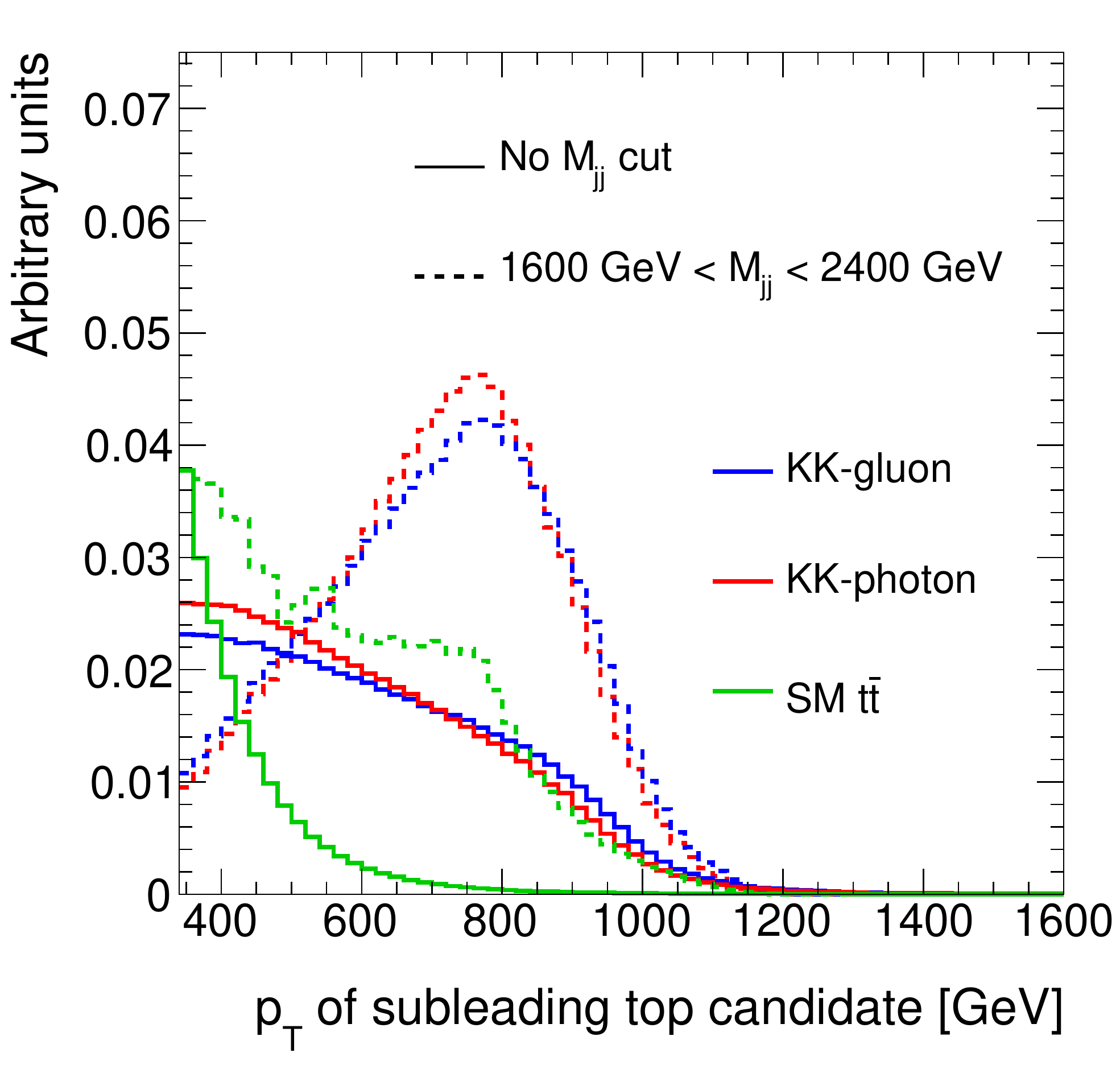} }%
}
\caption{
Transverse momentum of the leading (a) and subleading (b) top-jet candidates as reconstructed using the Cambridge-Aachen algorithm with $R=0.8$. The KK-gluon, KK-photon and SM $t\bar{t}$ events are shown in blue, red and green, respectively. Two distributions are shown for each sample, with no specific cut on the dijet invariant mass (solid lines) and after a dijet invariant mass cut of $1.6 \leq M_{\rm jj} < 2.4$~TeV (dashed lines).\label{fig:jetpt}
 }
\end{figure}

The Cambridge-Aachen (CA) algorithm \cite{ca_algo} as implemented in Fastjet \cite{fastjet} is used to reconstruct jets using all stable final state interacting particles\footnote{i.e. excluding neutrinos} with $|\eta|<4.9$, which is the coverage of the ATLAS and CMS calorimeters. The default CA distance parameter is $R=0.8$, unless otherwise stated, and full four momentum recombination is used. Jets are initially kept for further analysis if they have $p_{\rm T}^{}>350$~GeV.  Figure \ref{fig:jetpt} shows the transverse momentum of the leading and subleading jets. Following a dijet mass requirement, $1.6 \leq M_{\rm jj} < 2.4$~TeV, the transverse momentum of the leading/subleading jets from heavy gluon and heavy photon resonance decays are very similar. Any differences in top-tagging efficiency will therefore not be a consequence of different kinematic distributions. 

\subsection{Dependence of top-tagging efficiency on event colour structure}

We define the single and double tagging efficiencies, $\epsilon_{1}$ and $\epsilon_{2}$, as
\begin{equation}
\epsilon_{1} = \frac{N_{\rm 1t}}{N_{\rm 2 j}} \quad {\rm and} \quad \epsilon_{2} = \frac{N_{\rm 2t}}{N_{\rm 2 j}},
\end{equation}
where $N_{\rm 2 j}$ is the number of events that contain two CA jets with $p_{\rm T}>500$~GeV and $|\eta|<2.5$.
$N_{\rm 1t}$ is the subset of these dijet events for which the leading jet is identified as a top-candidate using one of the algorithms presented in Sec.~\ref{sec:toptag}. $N_{\rm 2t}$ is the subset of the dijet events for which the leading and subleading jets are identified as a top-candidates.

\begin{figure}[t]%
\centering
\mbox{
\subfigure[]{\includegraphics[width=0.5\textwidth]{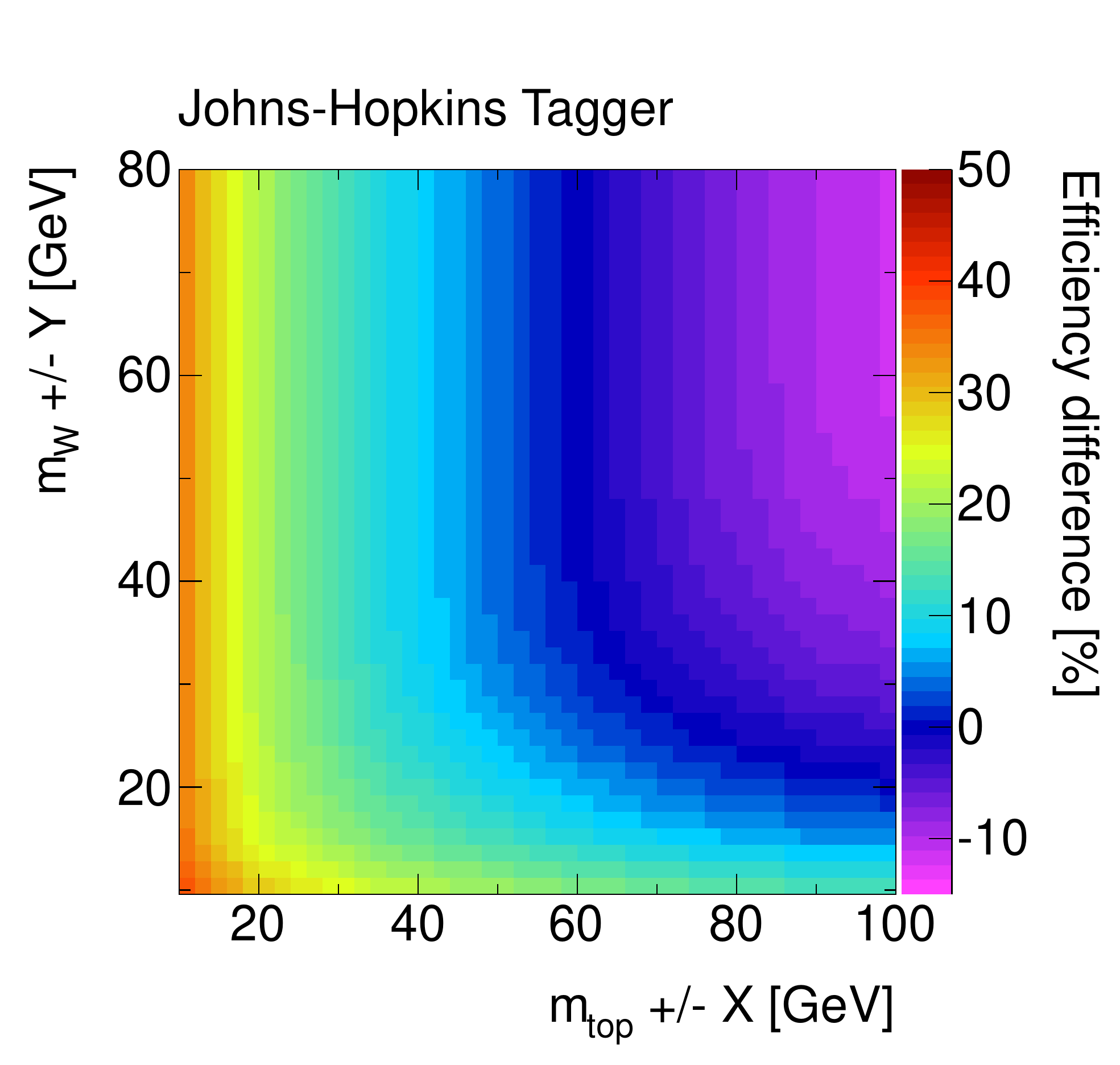}}\quad
\subfigure[]{\includegraphics[width=0.5\textwidth]{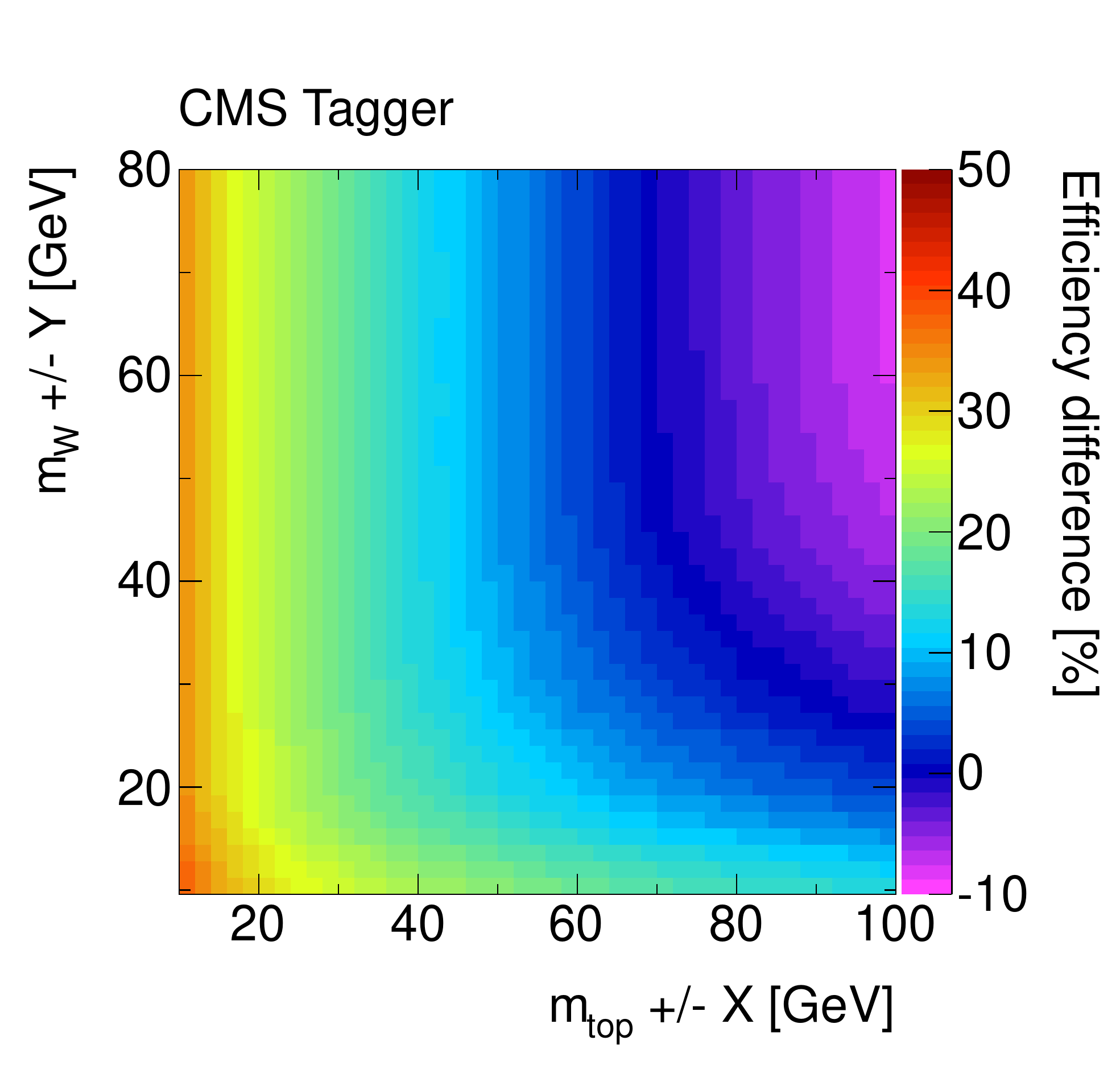} }
}
\mbox{
\subfigure[]{\includegraphics[width=0.5\textwidth]{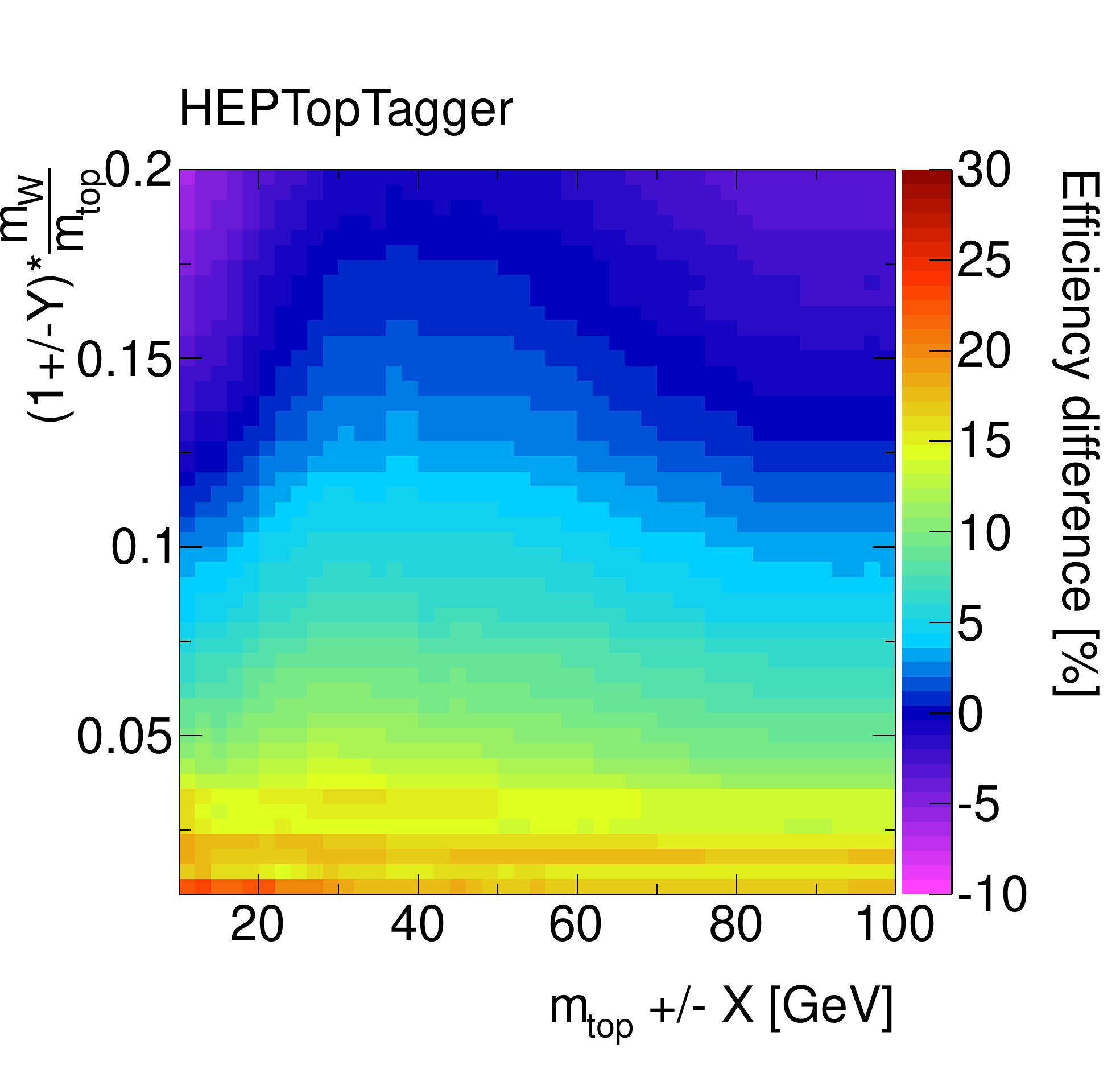} }\quad
\subfigure[]{\includegraphics[width=0.5\textwidth]{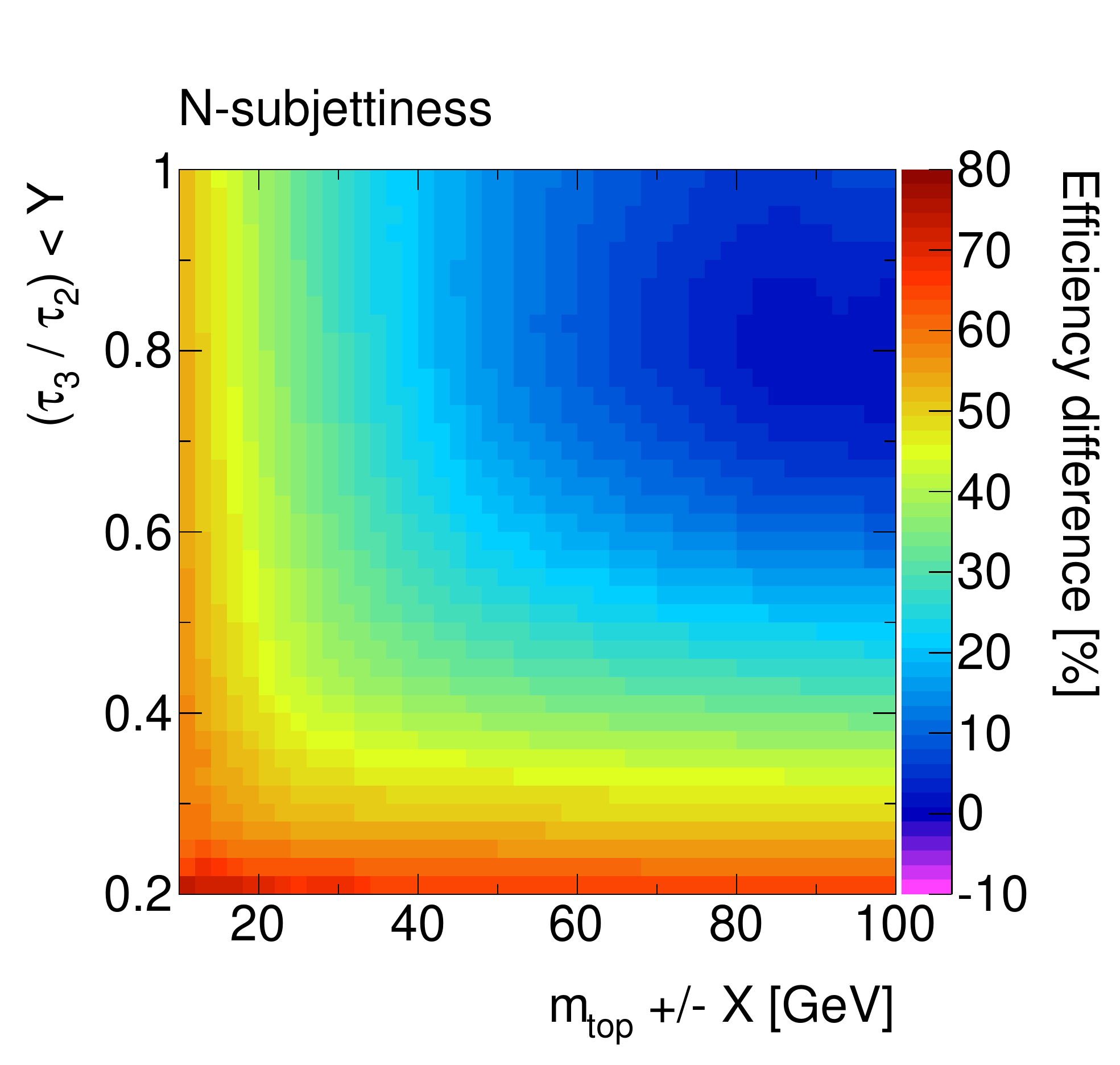} }
}
\caption{Efficiency difference for double top-tagging efficiencies for the (a)  \jh{}, (b) \cms{}, (c) \heptt{} and (d) \nsub{} algorithms. The difference of the efficiencies is presented as a function of the cuts applied to the algorithm input parameters. For example, $m_{\rm top} \pm X$ refers to the size of the window for which the top mass reconstructed from jet constituents is required to satisfy. Details pertaining to the top mass reconstruction and other algorithm parameters are given in Sec.~\ref{sec:toptag}. 
\label{fig:effvsparams_gluon}}
\end{figure}

The top-tagging efficiency depends on an educated guess as to the jet substructure that resembles a top quark decay. 
Figure \ref{fig:effvsparams_gluon} shows the relative difference between the double tagging efficiency for KK gluon and KK photons as a function of two parameters for each algorithm. The algorithm parameters are discussed in Sec.~\ref{sec:toptag}. 
It is immediately apparent that there is a difference between the double tagging efficiencies for colour octet and colour singlet resonances. This difference varies between 0\% and 75\%, depending on the algorithm and the parameter choice.  We find the largest efficiency differences occur when a high background rejection rate is targeted, i.e. for small mass windows and/or tight parameter cuts. This efficiency difference is directly due to the different colour structure in the events: There is more radiation present in events containing heavy photons, as observed in previous work \cite{Sung:2009iq,Ask:2011zs}. This difference in quark/gluon activity in each type of event means that there is more chance of radiation into the jets and hence more chance of failing various subjet requirements. For each algorithm, parameters can be chosen that yield the same average tagging efficiency for a color octet and a color singlet resonance. However, these parameters tend to correspond to very loose cuts on the subjets and would have poor background rejection.

Signal tagging efficiencies are usually quantified as a function of the mis-tag probability for the light quark and gluon background, so that the performance of each algorithm can be compared at similar working points. To do this, the two-dimensional tagging efficiencies for each signal and the background are projected onto one dimensional slices, allowing efficiency/mis-tag curves to be constructed. The choice of projection is somewhat arbitrary and so we choose three projections for each algorithm. For the \jh{} and \cms{} algorithms, efficiency/mis-tag curves are constructed by varying (i) only the top mass window (corresponding to a horizontal slice at $m_W \pm 30$~GeV), (ii) only the $W$ mass window (corresponding to a  slice at $m_t \pm 50$~GeV), or (iii) both the top and the $W$ mass windows simulataneously such that the $W$ mass window is approximately 80\% of the size of the top mass winsow (corresponding to a slice across the diagonal of Figure \ref{fig:effvsparams_gluon} (a)). For the \nsub{} and \heptt{} algorithms, we vary the $\tau_3/\tau_2$ cut and $m_W^{}/m_t^{}$ window, rather than the $W$ mass window. All other parameters of the algorithms are fixed to the default values. 

\begin{figure}[t]%
\centering
\mbox{
\subfigure[]{\includegraphics[width=0.49\textwidth]{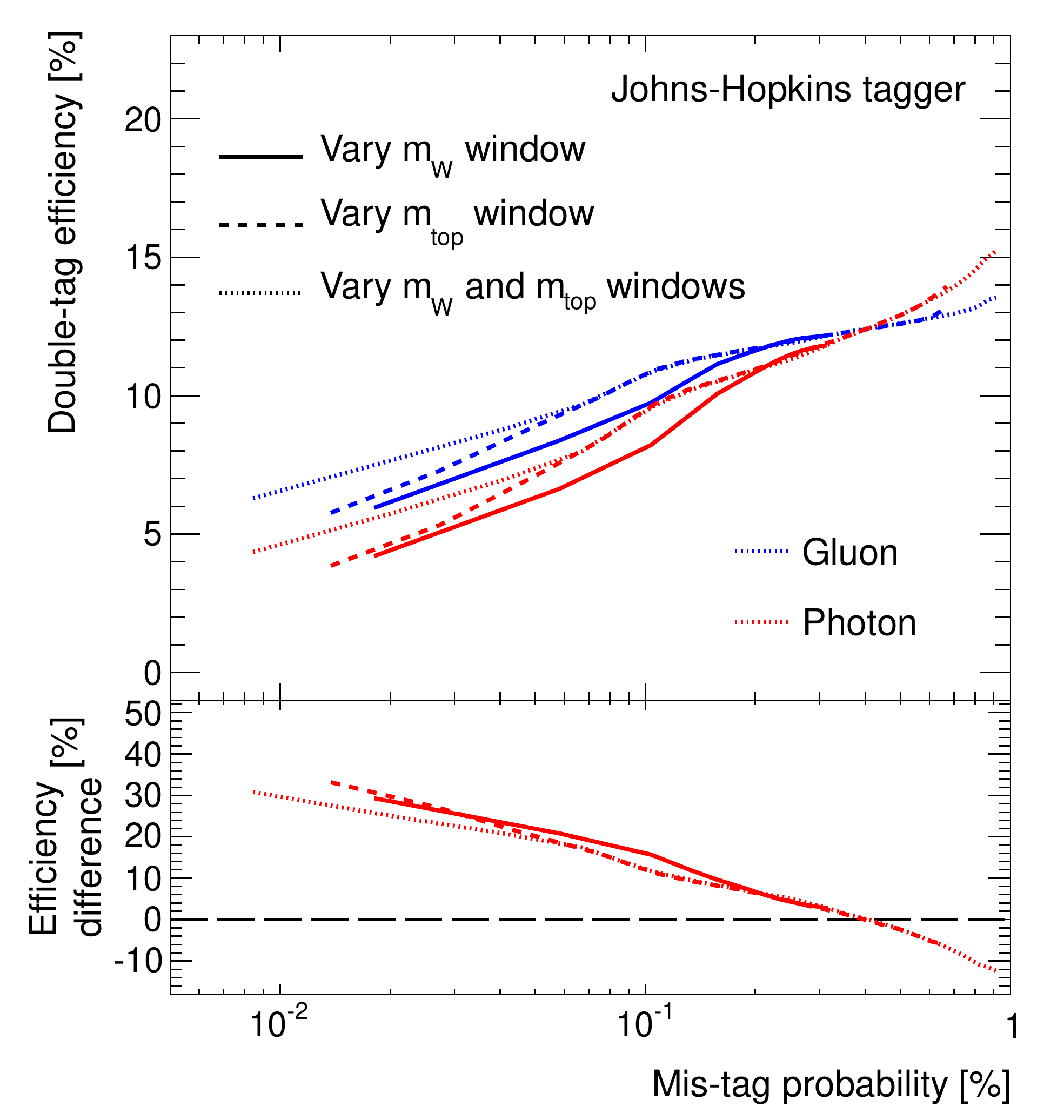}}\quad
\subfigure[]{\includegraphics[width=0.49\textwidth]{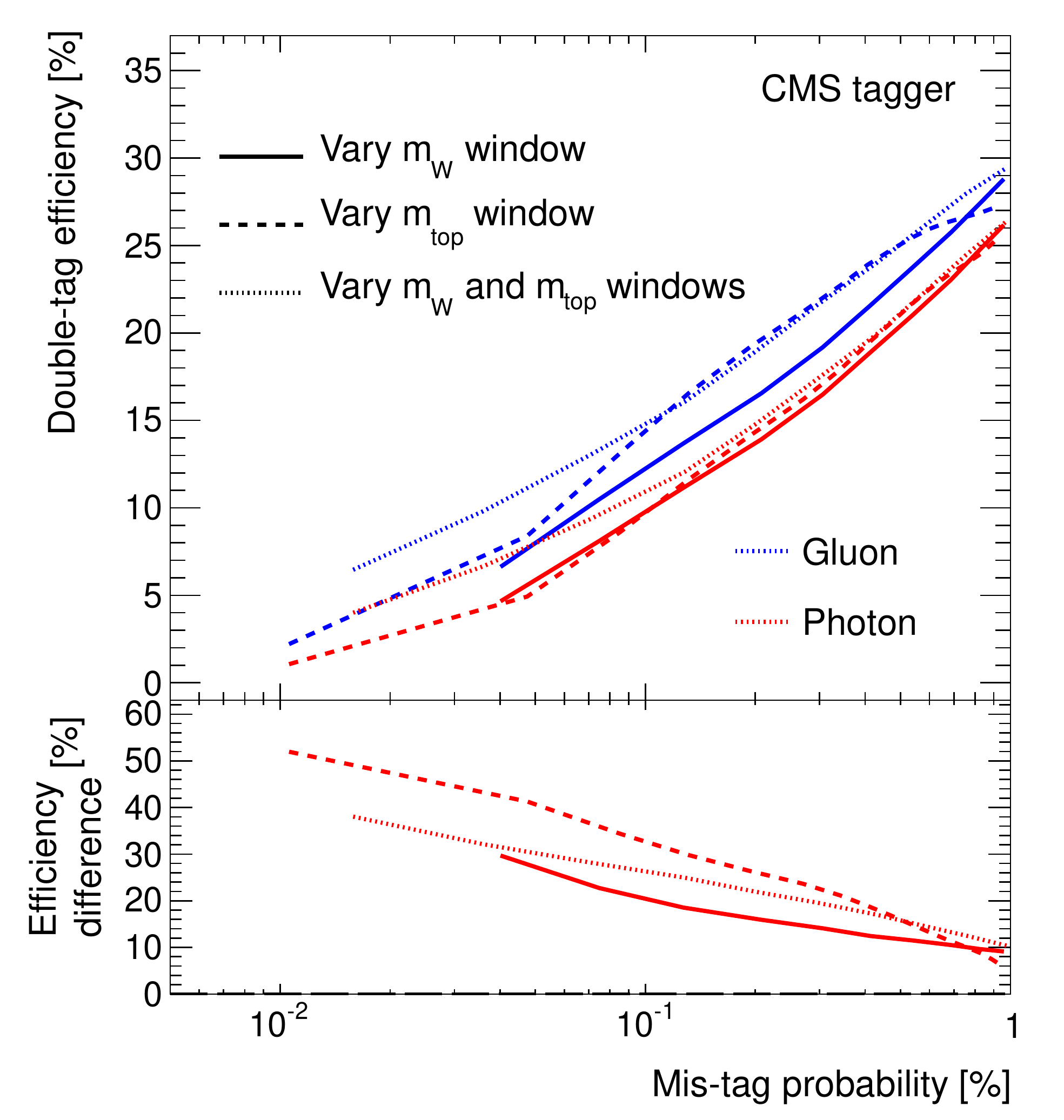} }
}
\mbox{
\subfigure[]{\includegraphics[width=0.49\textwidth]{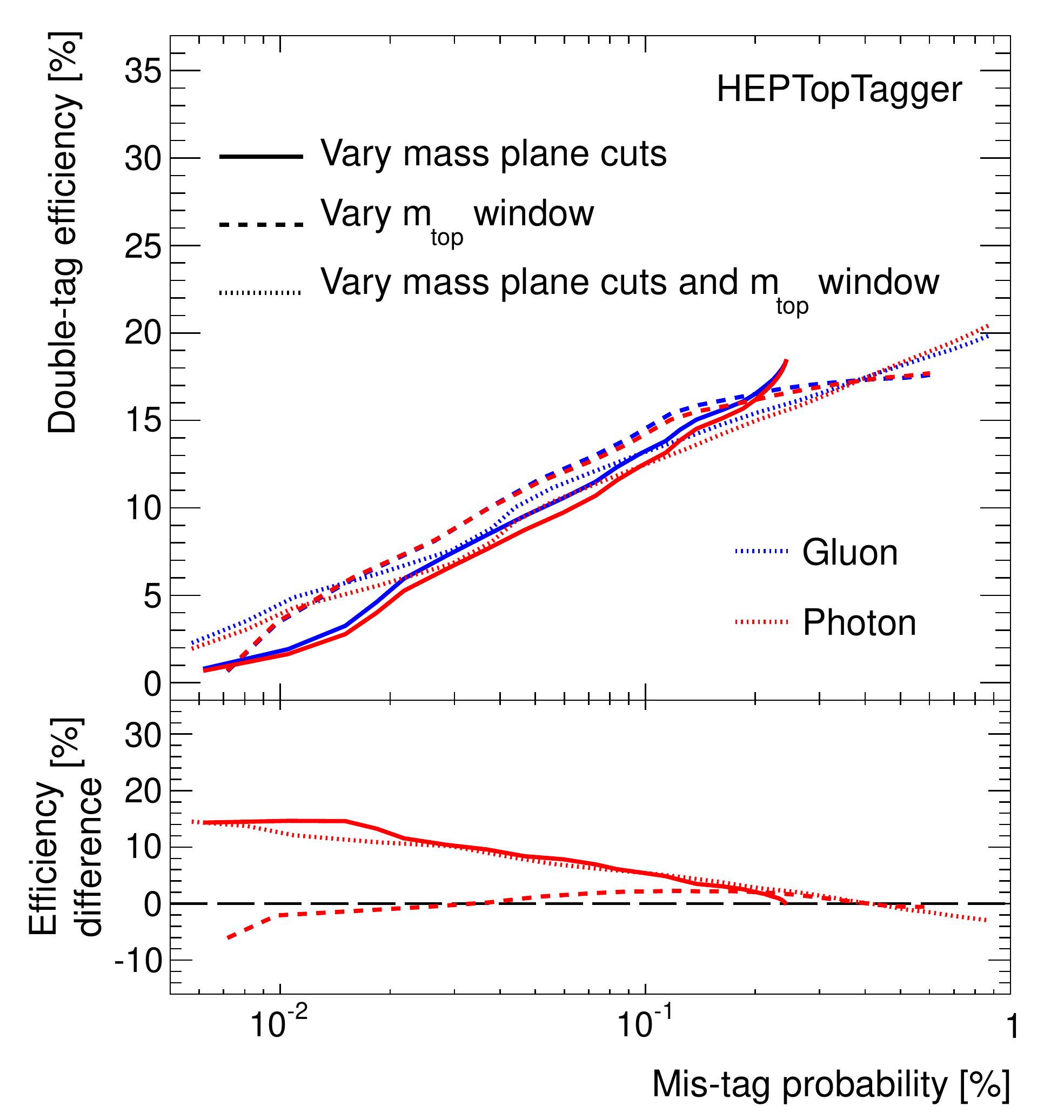} }\quad
\subfigure[]{\includegraphics[width=0.49\textwidth]{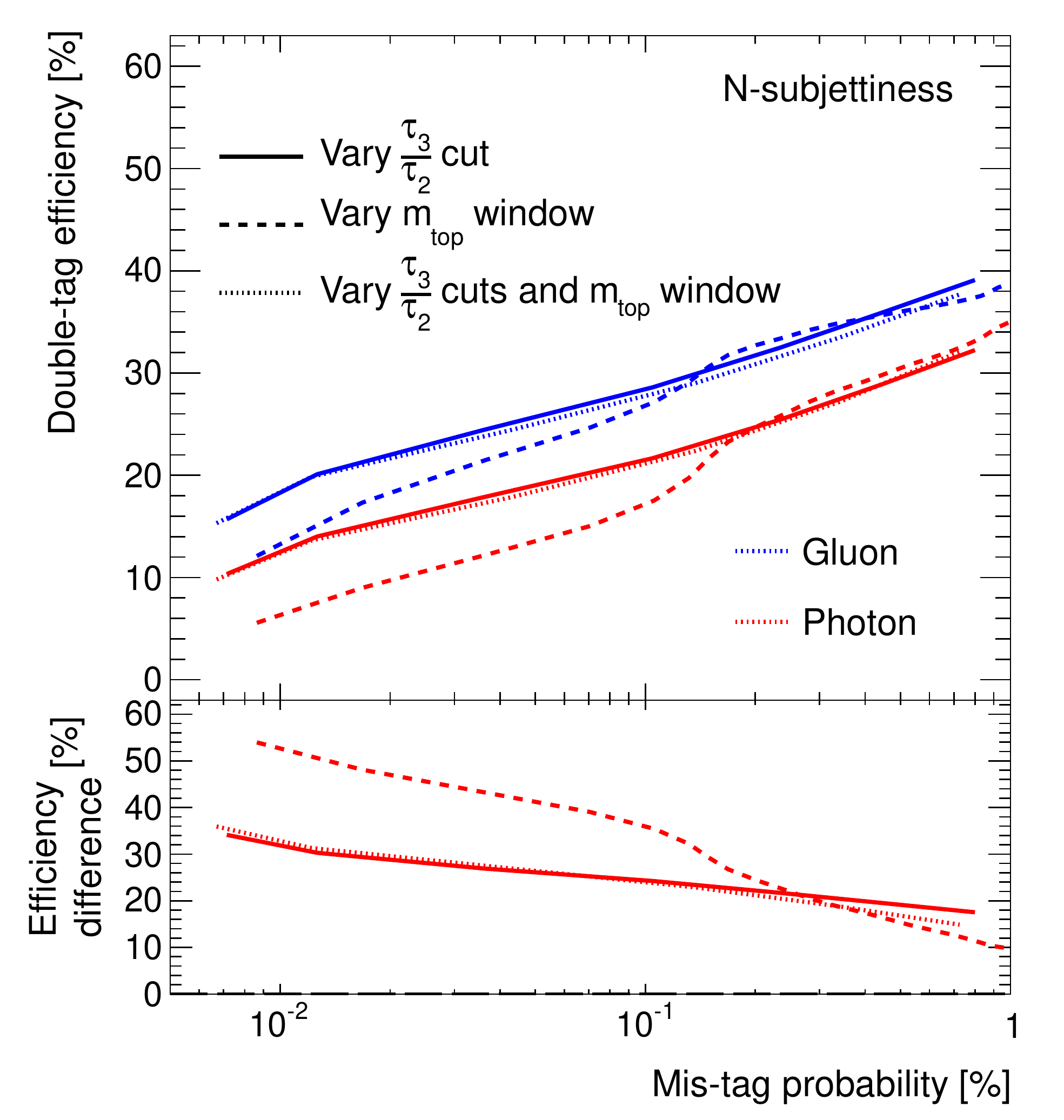} }
}
\caption{Efficiency of correctly tagging the leading and subleading jets from the decay of a gluon resonance (blue) and photon resonance (red), using the (a) \jh{}, (b) \cms{}, (c) \heptt{} and (d) \nsub{} algorithms. The efficiency is presented as a function of the double mis-tag probability, which is the probability of misidentifying two light jets as top-candidates. The efficiency and mis-tag probabilities are estimated by varying the cuts on the algorithm parameters. The ratio plots show the relative signal efficiencies ($\epsilon_{2}^{\rm gluon}/ \epsilon_{2}^{\rm photon} -1$) as a function of the mis-tag probability for each algorithm and parameter variation.
\label{fig:effvsmistag_2tags}}
\end{figure}

Figure \ref{fig:effvsmistag_2tags} shows the double top-tagging efficiency for heavy gluons and heavy photons as a function of the mis-tag probability for each top-tagging algorithm. The efficiency versus mis-tag curves for each signal are highly dependent on the choice of projection. As the signal efficiency at a given mis-tag working point is dependent on the choice of parameters, the algorithm parameters need to be optimised in any given search. We do not discuss the absolute values of the signal efficiencies further in this article, but instead focus on the tagging efficiency {\it differences} for KK gluons and KK photons. In this respect, we find that the \jh{} tagger does not show a strong dependence on the parameter varied, producing similar large efficiency differences if the mis-tag probability is required to be small. This is not true for the other taggers. For the \nsub{} and \cms{} taggers, the biggest tagging differences are found by varying the top mass window only. Conversely, the \heptt{} is almost completely insensitive to the variation of the top mass window, being more dependent on the mass plane window variation, which can result in a $15 \%$ double tagging efficiency difference between KK gluons and KK photons. This implies that searches could be designed to minimise the model-dependency for any given mis-tag working point by choosing algorithm parameters carefully. It is worth noting that, in searches for heavy resonances, working points corresponding to low  mis-tag probabilities $(\leq 1\%)$ are favoured to overcome the large QCD backgrounds. However, this is exactly the region that yields the largest tagging efficiency differences between KK gluons and KK photons, typically ranging from 15\% to 50\%.

\begin{figure}[t]%
\centering
\mbox{
\subfigure[]{\includegraphics[width=0.49\textwidth]{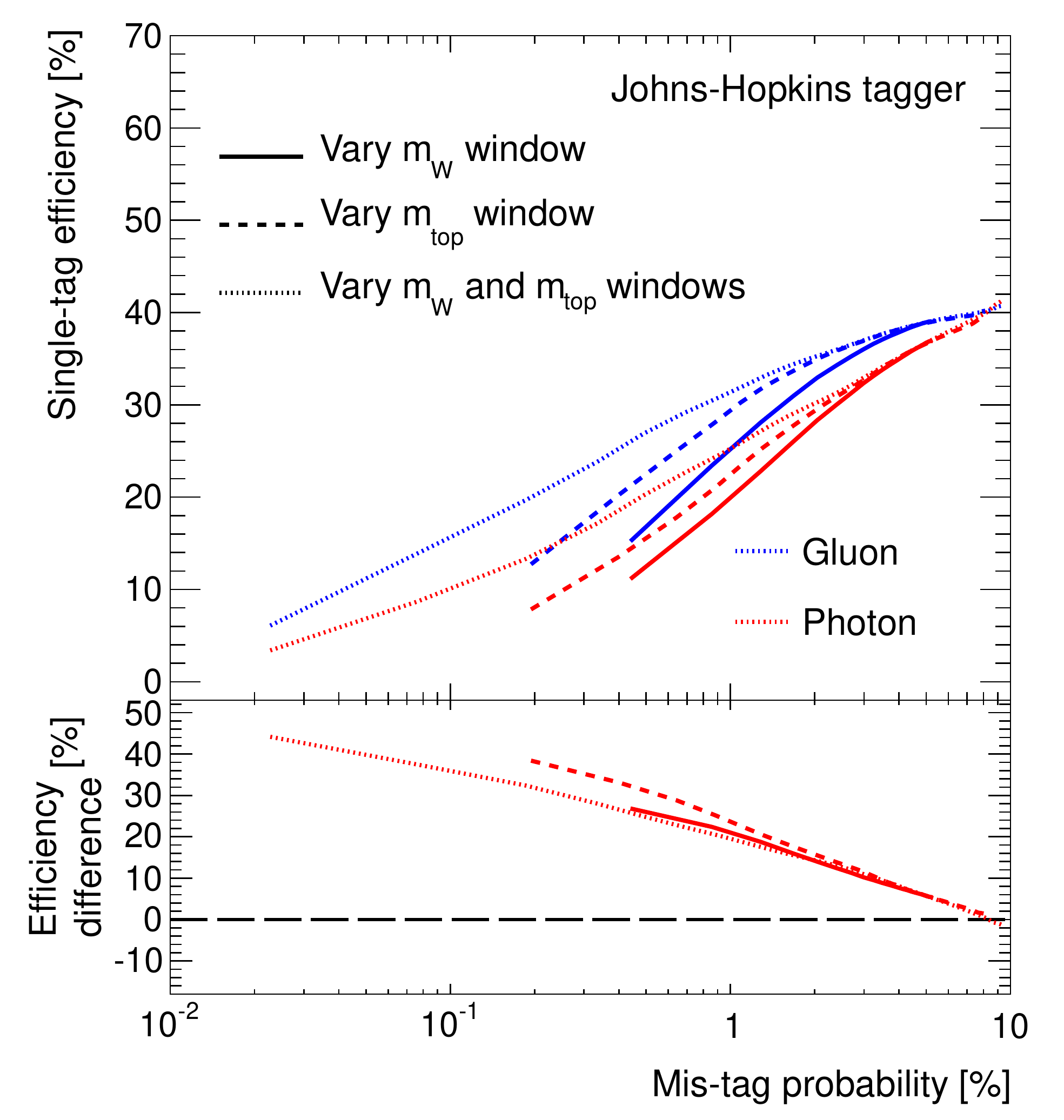}}\quad
\subfigure[]{\includegraphics[width=0.49\textwidth]{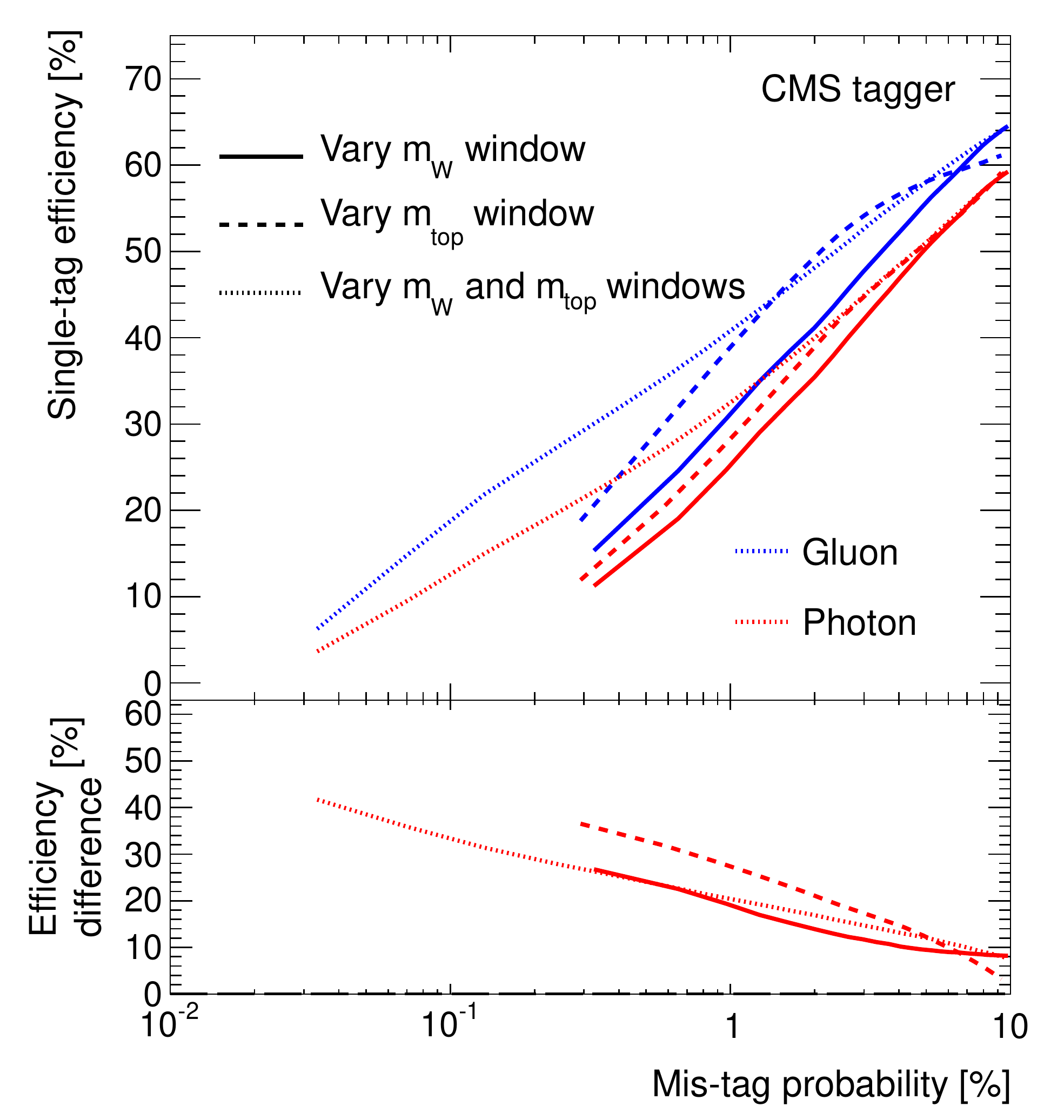} }
}
\mbox{
\subfigure[]{\includegraphics[width=0.49\textwidth]{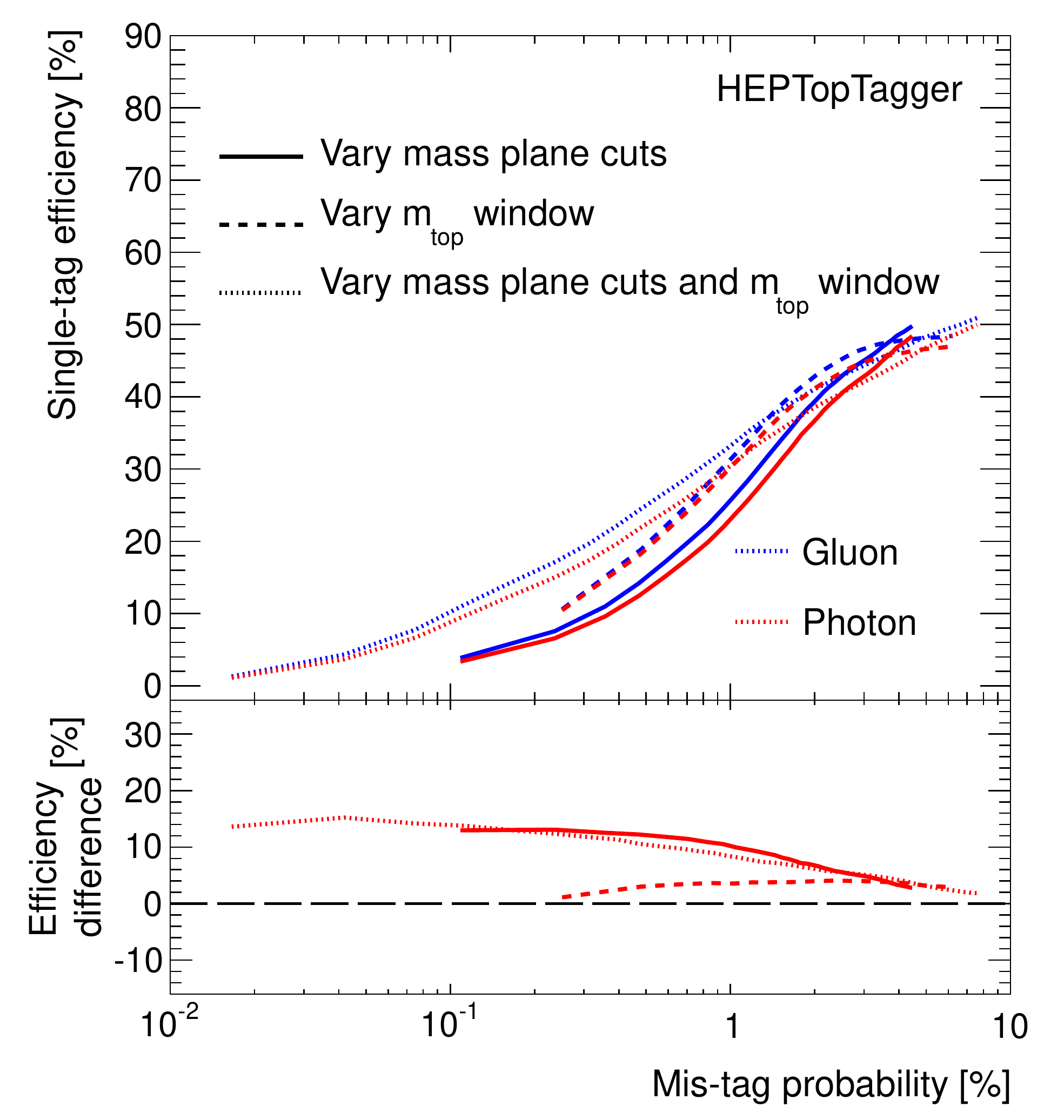} }\quad
\subfigure[]{\includegraphics[width=0.49\textwidth]{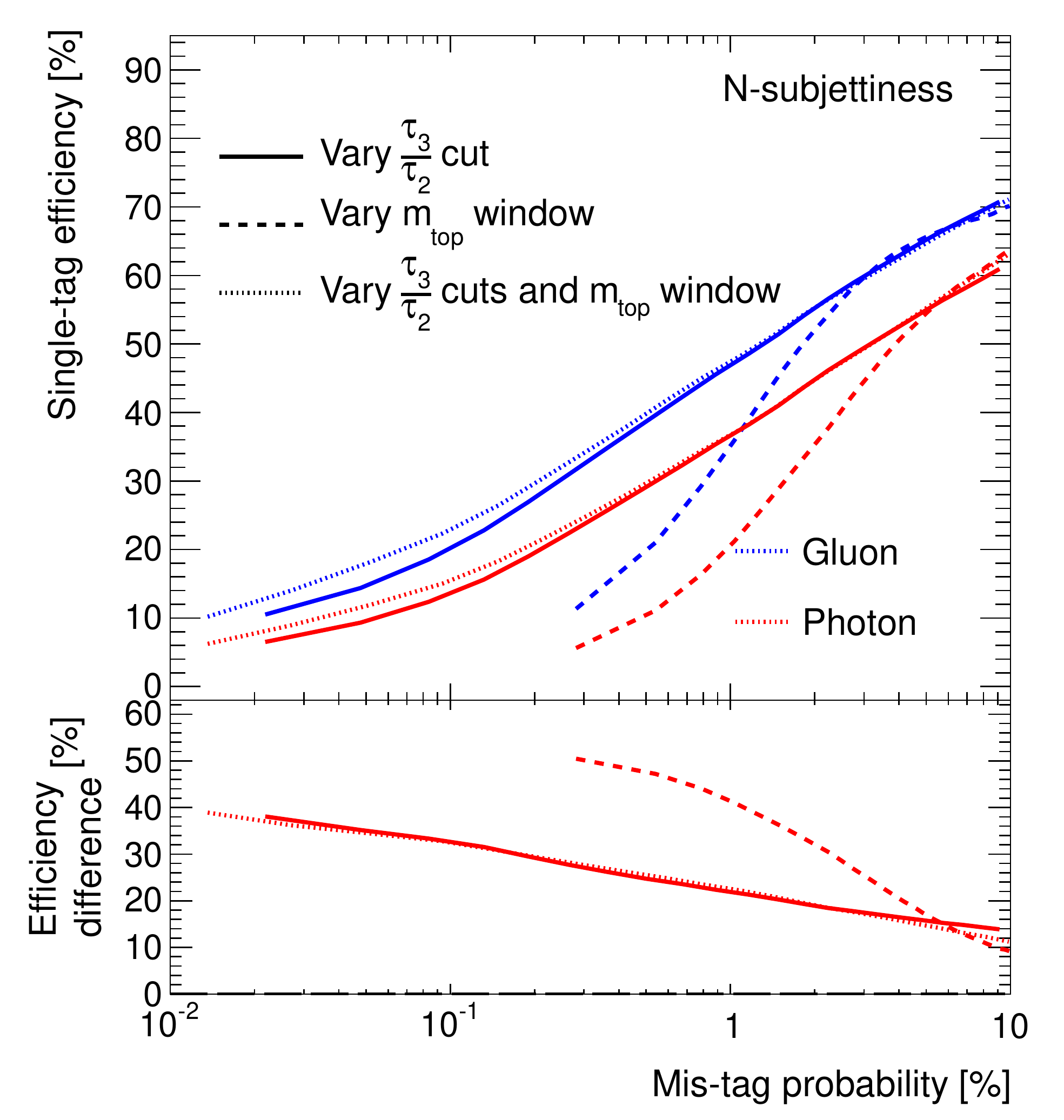} }
}
\caption{
Efficiency of correctly tagging the leading jet from the decay of a gluon resonance (blue) and photon resonance (red), using the  (a) \jh{}, (b) \cms{}, (c) \heptt{} and (d) \nsub{} algorithms. The efficiency is presented as a function of the single mis-tag probability, which is the probability of misidentifying the leading jet in a light-quark or gluon sample as the top-candidate. The efficiency and mis-tag probabilities are estimated by varying the cuts on the algorithm parameters. The ratio plots show the relative signal efficiencies ($\epsilon_{1}^{\rm gluon}/ \epsilon_{1}^{\rm photon}-1$) as a function of the CA distance parameter for each top-tagging algorithm.
\label{fig:effvsmistag_1tag}}
\end{figure}

The difference in tagging efficiencies is also present when tagging just the leading jet in the event, as shown in Figure \ref{fig:effvsmistag_1tag}. This demonstrates that the different colour structure is affecting both jets, not just the subleading jet. This is particularly important for searches in a semi-leptonic channel, in which only one jet is tagged as a top jet. It also implies that a tight/loose cut combination on the leading/subleading jets would not reduce the observed model dependency.

\begin{figure}[t]%
\centering
\mbox{
\subfigure[]{\includegraphics[width=0.49\textwidth]{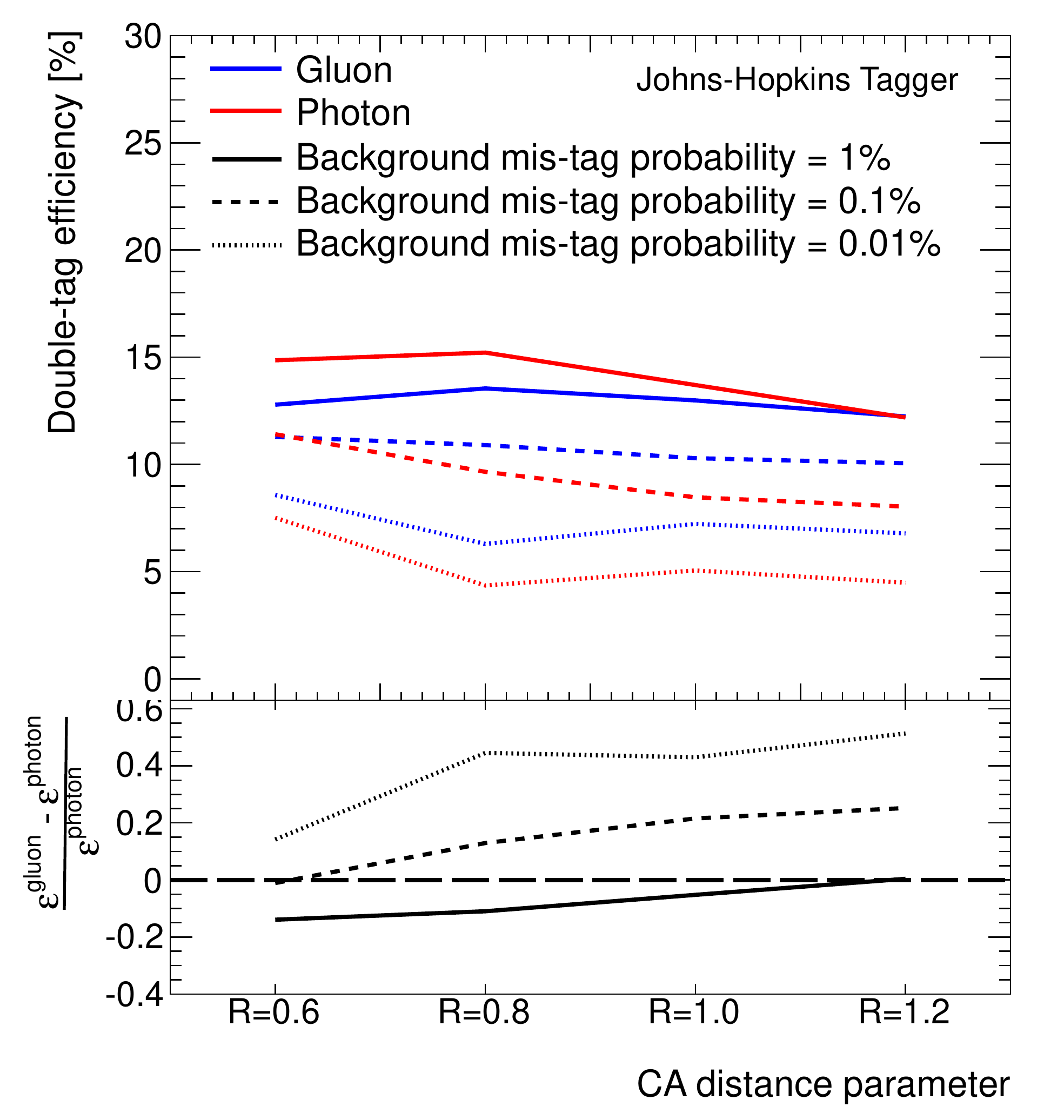}}\quad
\subfigure[]{\includegraphics[width=0.49\textwidth]{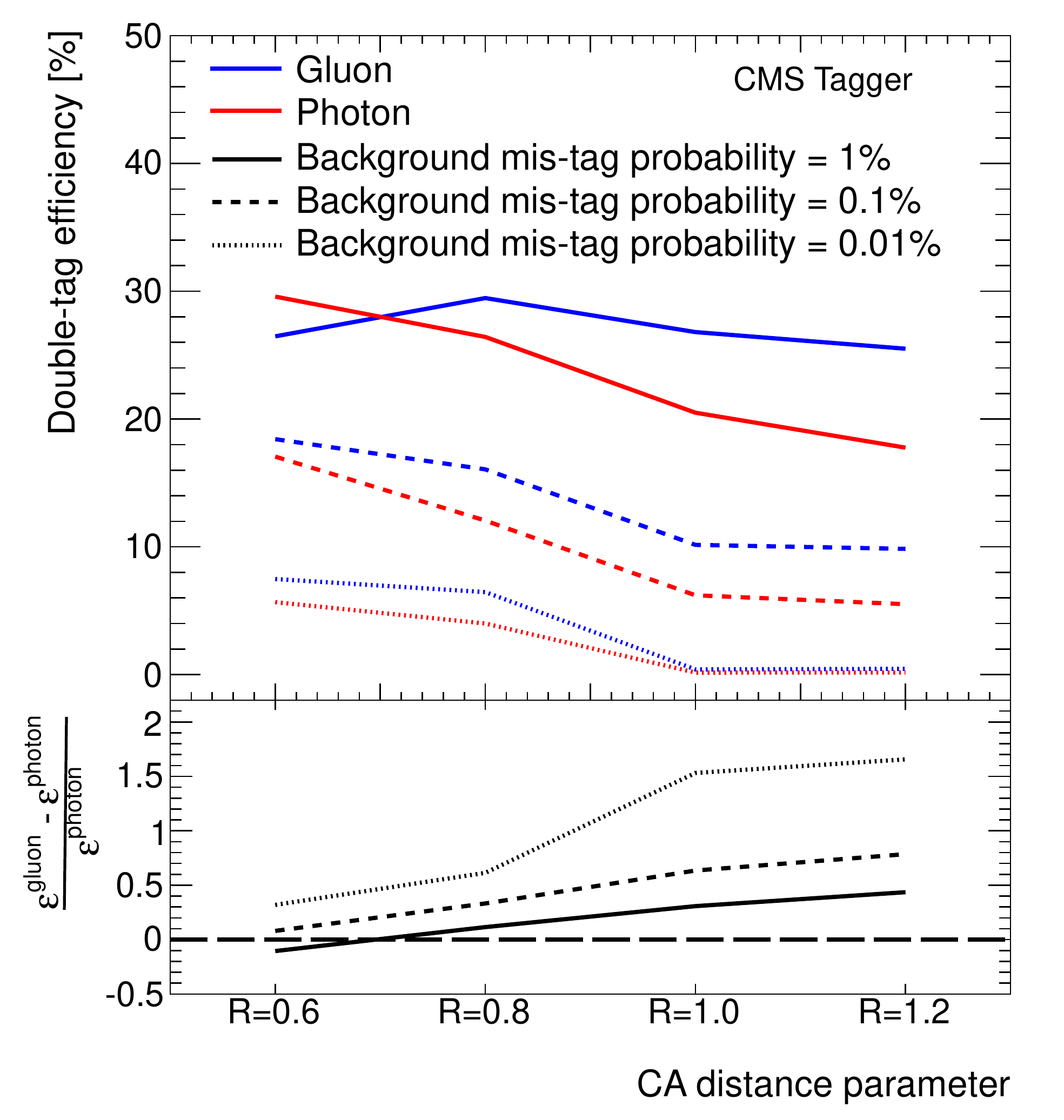} }
}
\mbox{
\subfigure[]{\includegraphics[width=0.49\textwidth]{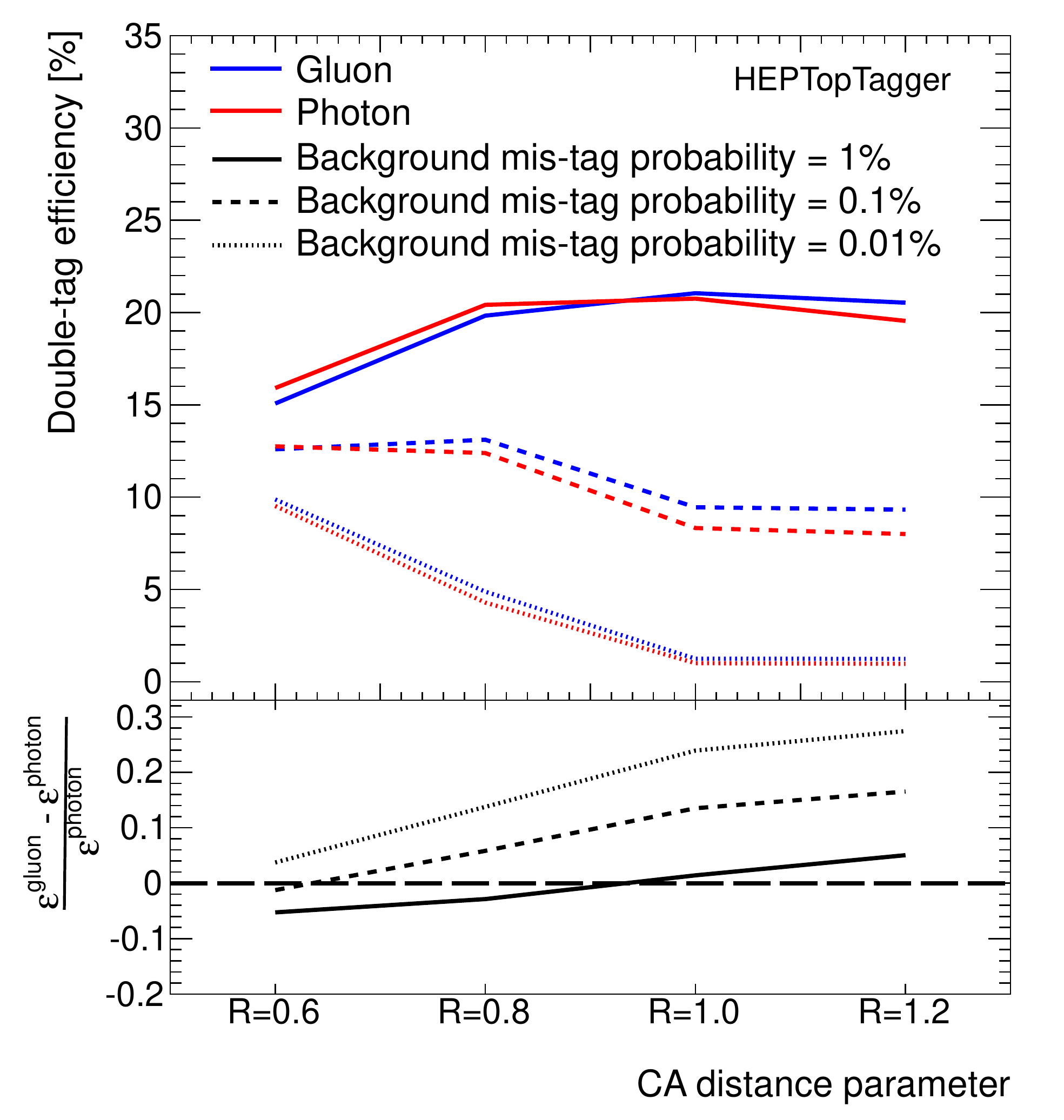} }\quad
\subfigure[]{\includegraphics[width=0.49\textwidth]{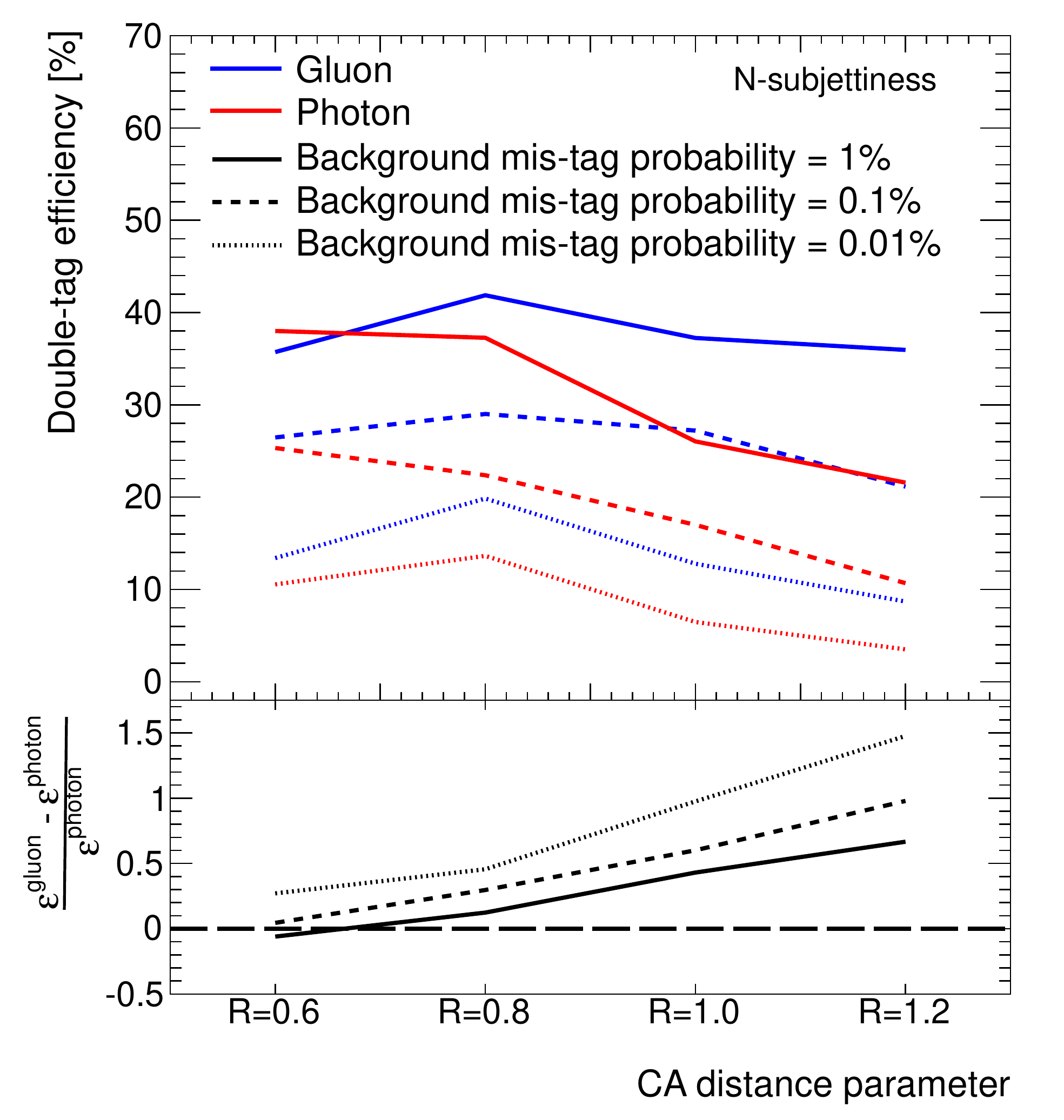} }
}
\caption{Efficiency of correctly tagging the leading and subleading jets from the decay of a gluon resonance (blue) and photon resonance (red), using the (a) Johns-Hopkins, (b) CMS, (c) \heptt{} and (d) N-subjettiness algorithms. The efficiency is presented as a function of the CA distance parameter used to reconstruct the jets. The efficiency and mis-tag probabilities are estimated by varying the cuts on the algorithm parameters. The ratio plots show the relative signal efficiencies ($\epsilon_{2}^{\rm gluon}/ \epsilon_{2}^{\rm photon} -1$) as a function of the mis-tag probability for each algorithm and parameter variation.
\label{fig:effvsparams_vs_R}}
\end{figure}

The size of the fat jet is a crucial parameter for any subjet analysis. In Figure \ref{fig:effvsparams_vs_R}, we examine the difference in the double tagging efficiency between heavy gluons and heavy photons as a function of the CA distance parameter, $R$. We show the tagging efficiency at each value of $R$ for three efficiency working points defined by the background mis-tag probabilities $(0.01\%,0.1\%,1\%)$. The algorithm parameters chosen to obtain those working points are presented in Table \ref{tab:cutvalues}. In general, the efficiency increases slightly as $R$ decreases. This is somewhat counter intuitive because a larger jet is more likely to capture all top decay products. However, in this study, the top quarks are typically boosted enough for their decay products to be captured by a jet size of $R \simeq 0.8$ and increasing the jet size simply allows more soft radiation to enter the jet. Finally, although the top-tagging efficiency only slightly decreases for larger $R$, the difference in tagging efficiencies between the KK gluon and KK photon increases rapidly. Optimising the jet size as a function of the top quark transverse momentum has already been suggested in the literature \cite{variableR}. These results imply that the minimal possible value of $R$ should be chosen, in order to minimise the dependence on the colour structure.

\begin{sidewaystable}
\begin{center}
\begin{tabular}{l | ccc | ccc | ccc | ccc}
\multicolumn{1}{c}{\textbf{JH Tagger}} & \multicolumn{3}{c}{R = 0.6} & \multicolumn{3}{c}{R = 0.8} & \multicolumn{3}{c}{R = 1.0} & \multicolumn{3}{c}{R = 1.2} \\
\midrule
Mis-tag rate [$\%$] & 1 & 0.1 & 0.01 & 1 & 0.1 & 0.01 & 1 & 0.1 & 0.01 & 1 & 0.1 & 0.01 \\
$m_{t} \pm X$ [GeV] & 97.0 & 41.0 & 21.0 & 97.0 & 33.0 & 15.0 & 97.0 & 37.0 & 17.0 & 97.0 & 39.0 & 19.0 \\
$m_{W} \pm Y$ [GeV] & 77.6 & 32.8 & 16.8 & 77.6 & 26.4 & 12.0 & 77.6 & 29.6 & 13.6 & 77.6 & 31.2 & 15.2 \\
\midrule
\multicolumn{13}{l}{} \\
\multicolumn{1}{c}{\textbf{CMS Tagger}} & \multicolumn{3}{c}{R = 0.6} & \multicolumn{3}{c}{R = 0.8} & \multicolumn{3}{c}{R = 1.0} & \multicolumn{3}{c}{R = 1.2} \\
\midrule
Mis-tag rate [$\%$] & 1 & 0.1 & 0.01 & 1 & 0.1 & 0.01 & 1 & 0.1 & 0.01 & 1 & 0.1 & 0.01 \\
$m_{t} \pm X$ [GeV] & 85.0 & 37.0 & 17.0 & 47.0 & 21.0 & 9.0 & 39.0 & 17.0 & 5.0 & 39.0 & 15.0 & 5.0 \\
$m_{W} \pm Y$ [GeV] & 68.0 & 29.6 & 13.6 & 37.6 & 16.8 & 7.2 & 31.2 & 13.6 & 4.0 & 31.2 & 12.0 & 4.0 \\
\midrule
\multicolumn{13}{l}{} \\
\multicolumn{1}{c}{\textbf{HEPT\textsc{op}T\textsc{agger}}} & \multicolumn{3}{c}{R = 0.6} & \multicolumn{3}{c}{R = 0.8} & \multicolumn{3}{c}{R = 1.0} & \multicolumn{3}{c}{R = 1.2} \\
\midrule
Mis-tag rate [$\%$] & 1 & 0.1 & 0.01 & 1 & 0.1 & 0.01 & 1 & 0.1 & 0.01 & 1 & 0.1 & 0.01 \\
$m_{t} \pm X$ [GeV] & 97.0 & 71.0 & 51.0 & 97.0 & 51.0 & 25.0 & 95.0 & 35.0 & 11.0 & 91.0 & 37.0 & 13.0 \\
$(1 \pm Y) * m_{W} / m_{t}$ & 0.19 & 0.14 & 0.10 & 0.19 & 0.10 & 0.05 & 0.19 & 0.07 & 0.02 & 0.18 & 0.07 & 0.03 \\
\midrule
\multicolumn{13}{l}{} \\
\multicolumn{1}{c}{\textbf{N-subjettiness}} & \multicolumn{3}{c}{R = 0.6} & \multicolumn{3}{c}{R = 0.8} & \multicolumn{3}{c}{R = 1.0} & \multicolumn{3}{c}{R = 1.2} \\
\midrule
Mis-tag rate [$\%$] & 1 & 0.1 & 0.01 & 1 & 0.1 & 0.01 & 1 & 0.1 & 0.01 & 1 & 0.1 & 0.01 \\
$m_{t} \pm X$ [GeV] & 69.0 & 59.0 & 43.0 & 65.0 & 55.0 & 45.0 & 65.0 & 55.0 & 41.0 & 65.0 & 55.0 & 43.0 \\
$\tau_{3} / \tau_{2} < Y$ & 0.69 & 0.59 & 0.43 & 0.65 & 0.55 & 0.45 & 0.65 & 0.55 & 0.41 & 0.65 & 0.55 & 0.43 \\
\midrule
\end{tabular}
\end{center}
\caption{
Cut values required by each tagger to obtain the background mis-tag rates shown in Figure~\ref{fig:effvsparams_vs_R}.
}
\label{tab:cutvalues}
\end{sidewaystable}

\subsection{Impact on experimental analyses}
\label{results2}

The dependence of the top-tagging efficiency on the event colour structure raises a number of experimental issues relating to the use of top-tagging algorithms in the search for, and possible discovery of, new phenomena. 

In searches for heavy particles decaying to $t\bar{t}$, the sensitivity to the resonance's colour charge poses an experimental challenge in assessing the systematic uncertainties of the search. The relevant issues are clarified by scrutinising the recent CMS search for resonances decaying to top quark pairs \cite{cmssearch}. The single top-tagging efficiency for a colour singlet resonance was estimated to be 50\% for a mis-tag probability of 5\%, using the \cms{} algorithm ($R=0.8$, fat-jet $p_{\rm T}^{}>500$~GeV). The systematic uncertainty in the tagging efficiency was estimated to be 3\% (6\%) for a single (double) tag, by comparing the $W$-tagging efficiency in data to MC using a sample of moderately boosted SM $t\bar{t}$ events. The same methodology also yielded a scale (normalisation) factor of 0.97 for MC events, to recover the central tagging efficiencies observed in the data. These scale factors and systematic uncertainties were then used to normalise all signal and background samples.

\begin{figure}[t]%
\centering
\mbox{
\subfigure[]{\includegraphics[width=0.49\textwidth]{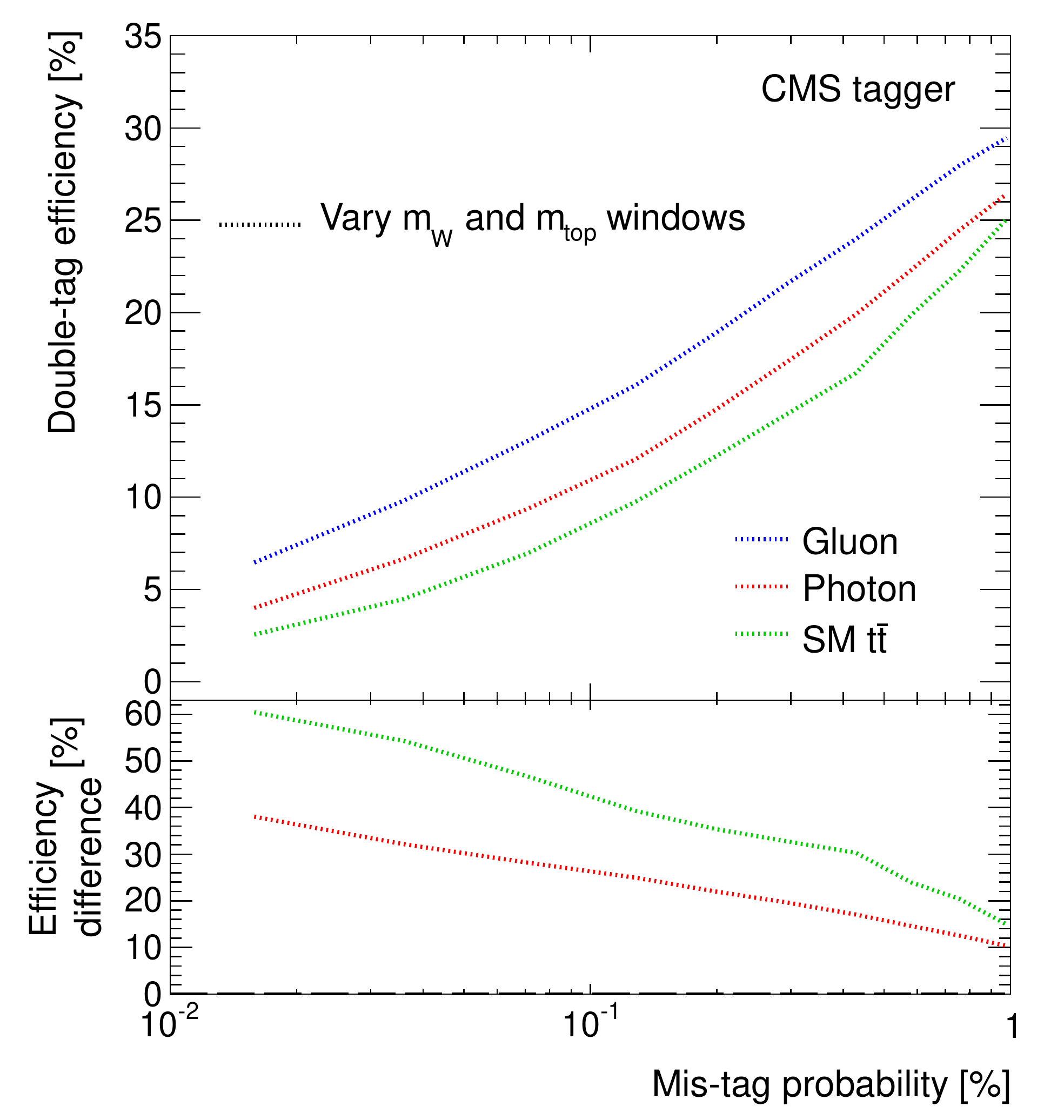}}\quad
\subfigure[]{\includegraphics[width=0.49\textwidth]{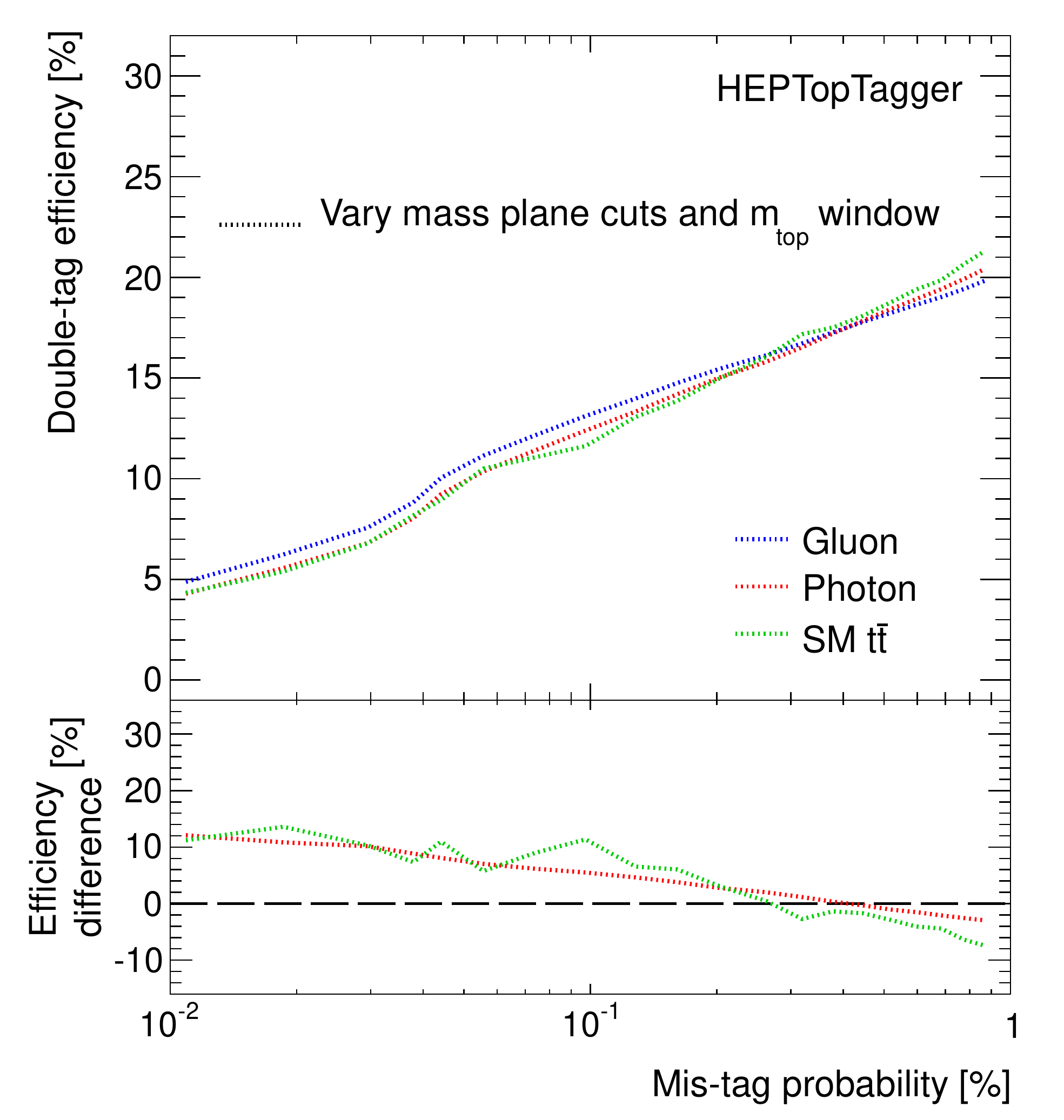} }
}
\caption{Double-tag efficiency and efficiency difference for the \cms{} tagger and the \heptt{}. We compare the KK-photon and the KK-gluon with the SM $t\bar{t}$ background.
\label{fig:effvsparams_withtop}}
\end{figure}

It is not immediately obvious that universal scale factors and systematic uncertainties can be used, given the colour flow dependency discussed in the previous section. The problem is that the data/MC agreements are observed in a control region, but the tagging performance can change dramatically between this control region and the signal regions. In the case of the CMS analysis, the efficiency scale factors and systematic uncertainties are determined from $W$-tagging in a sample of $t\bar{t}$ events and then assumed to be directly applicable to (i) top-tagging in the background MC samples and (ii) top-tagging in the signal MC samples. Both of these assumptions directly impact on the observed limit (although the dominant effect on the limit is usually from the background estimation). We note that the application of the scale factors and uncertainties to the background $t\bar{t}$ sample is not unreasonable because the colour flow in the control and background samples will be similar. However, the colour flow is very different between the control and signal regions, resulting in different tagging efficiencies as shown in Figure \ref{fig:effvsparams_withtop}; the difference between the double tagging efficiency for the SM $t\bar{t}$ events and the KK photon (gluon) is approximately 20\% (35\%) at the nominal mis-tag working point of 0.25\% ($\equiv 5\%^2$). Extrapolating from the control region to the signal region therefore requires much confidence in the MC generators. Finally, we note that the dependency on MC could be reduced in future analyses by optimising the choice of tagging algorithm and/or tagging parameters. For example, Figure \ref{fig:effvsparams_withtop} also shows that the \heptt{} has much smaller differences between control and signal regions.

The situation would become even more problematic if a heavy resonance was observed at the LHC. Without prior knowledge of the resonance colour, the cross-section$\times$branching-ratio could not be extracted in a model independent way. It is worth noting that methods to extract the colour of a heavy resonance have been discussed in the literature \cite{Gallicchio:2010sw,spanno_soper,Hook:2011cq,Ask:2011zs,Sung:2009iq}. Using one or more of these methods would be mandatory to reduce uncertainties related to the tagging efficiency.

The dependence of the tagging efficiency on the colour flow will impact upon other searches if boosted tagging algorithms are used for signal identification. We briefly highlight two other search channels of current interest. The first case is the search for scalar top (stop) pair production in which the (light) stops are produced in a boosted topology \cite{boostedstops}. In the worst case scenario, one or more of the stops are produced via a decay chain (e.g from gluino decays). In this case, the colour flow is very different to SM $t\bar{t}$ production and the determination of the signal tagging efficiency and associated systematic uncertainty will have to fully rely on the MC simulation. The second case is the search for a very broad resonance (i.e. with ~TeV width) decaying to top-quark pairs. In this scenario, distinguishing between the signal and background events is not possible and the background control regions will be contaminated by the signal. In this case, the MC will have to be fully relied upon to untangle the relative efficiency differences between signal and background.

\section{Summary and outlook}
\label{sec:outlook}



The impact of event colour structure on the performance of standard top-tagging algorithms was investigated by studying Monte Carlo simulations of colour singlet and colour octet resonances decaying to top-quark pairs. Large differences in top-tagging efficiency were observed due to the different colour charge of each resonance. Attempting to reduce the mis-tag probability, by tightening the algorithm cuts, leads to the largest differences in the top-tagging efficiency for colour singlet and colour octet resonances. The difference in tagging efficiency was also found to be largest for large values of the distance merging parameter ($R$) used to reconstruct the CA jet. The algorithm with the least overall sensitivity was the \heptt{}, which is the only algorithm with a built in jet grooming procedure to remove contamination from soft-radiation into the jet. Optimising the algorithm and choice of parameters would minimise this model dependency in future experimental searches.

The model dependency directly impacts on the searches that use boosted top-tagging algorithms. Furthermore, in the advent of a discovery, the model dependence will prevent a precise extraction of the cross section $\times$ branching ratio for the newly observed resonance. The problem, experimentally, is related to the lack of a suitable control region to determine the tagging efficiency for the signal events and the measurements may have to rely on the simulation provided by the MC event generators.

The observation that the tagging efficiency depends strongly on the events colour flow has wider relevance than the case presented in this article (searching for narrow resonances decaying to top quark pairs). For example, the search for the pair production of light scalar tops using boosted reconstruction techniques would also suffer from the lack of identifiable control regions for signal. Furthermore, the search for a broad resonance decaying to top-quark pairs would have a background control region contaminated by the signal and untangling the efficiency differences would be entirely dependent on the MC models.

\begin{acknowledgments}

The authors would like to thank Sarah Livermore and Mark Owen for interesing discussions. This work was funded in the UK by STFC.

\end{acknowledgments}

\section*{Appendix A}

\begin{figure}[t]%
\centering
\mbox{
\subfigure[]{\includegraphics[width=0.47\textwidth]{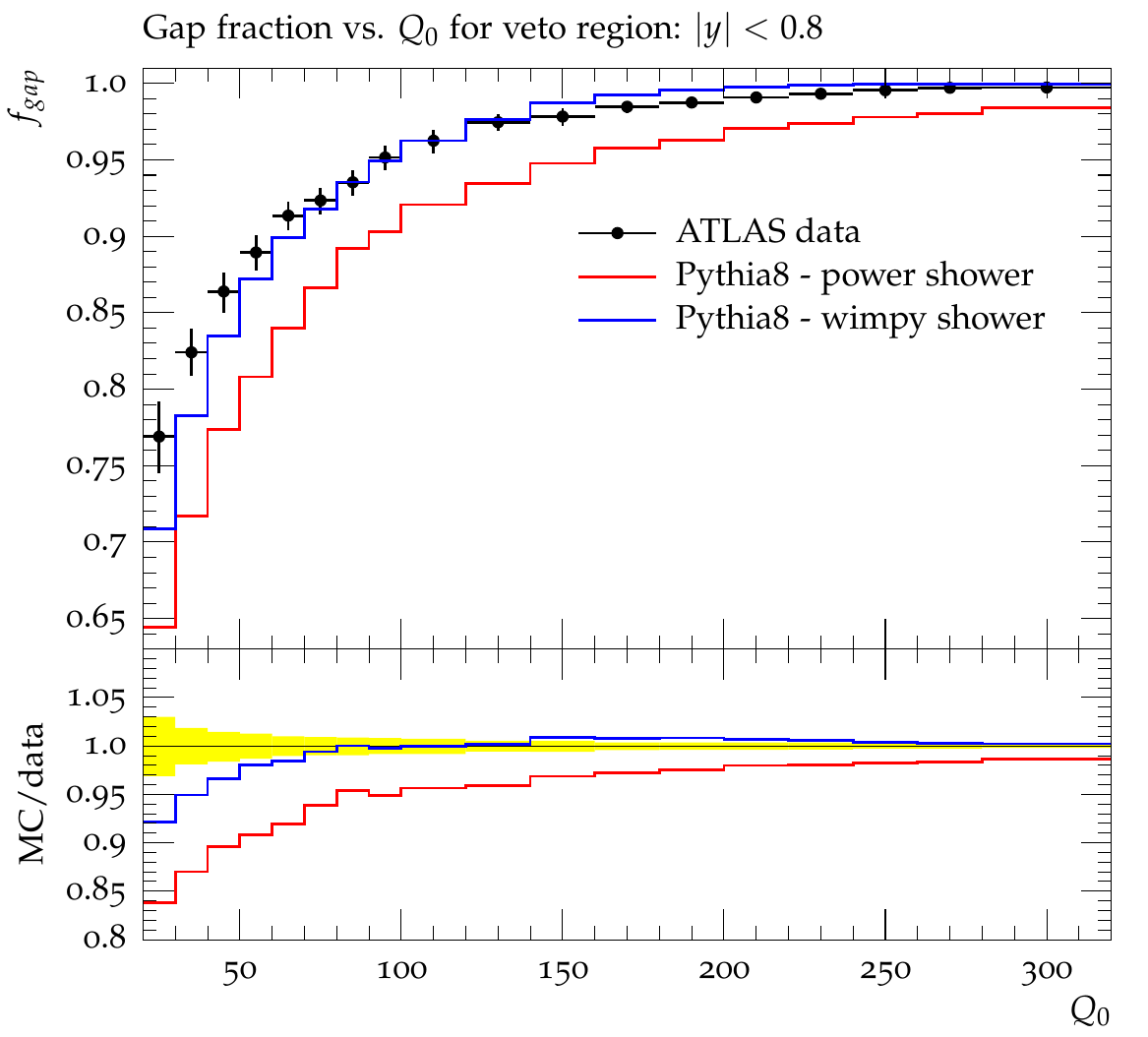}}\quad
\subfigure[]{\includegraphics[width=0.47\textwidth]{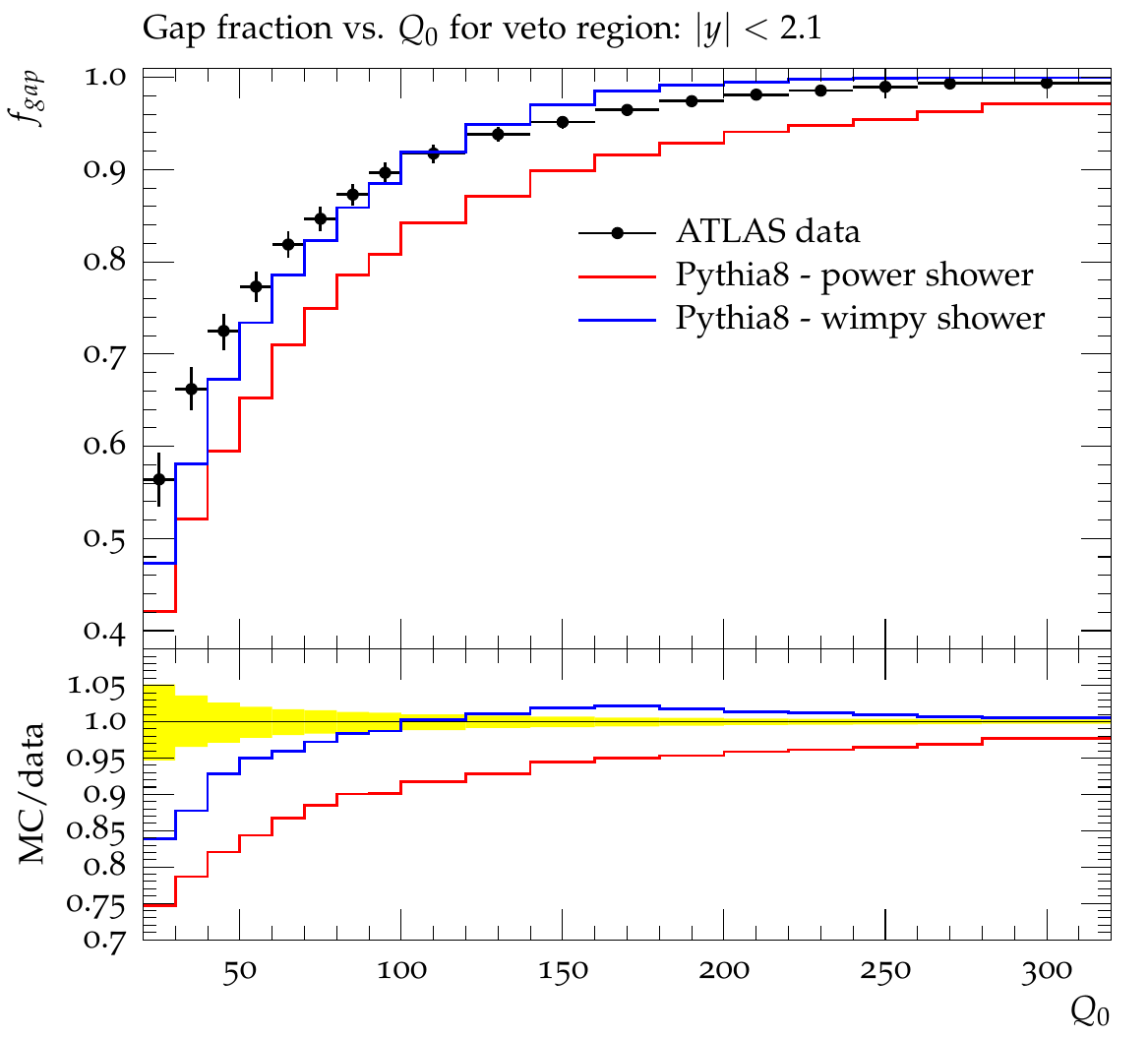}}
}
\caption{The fraction of $t\bar{t}$ events that do not contain an additional jet with $p_{\rm T} > Q_0$. The ATLAS data \cite{ATLAS:2012al} are compared to the predictions from {\sc pythia 8} using the so-called `power' (red) and `wimpy' (blue) parton shower options. The data clearly favour the wimpy shower, which was subsequently used for all the MC event generation in this study.\label{fig:atlaspower}}
\end{figure}

For  $t\bar{t}$ final states produced using the {\sc pythia 8} event generator, it is possible to choose two parton showering options; `wimpy' showers allow QCD radiation up to the factorisation scale, whereas power showers allow radiation up to the kinematic limit. We compared the results of using the two parton shower options to the ATLAS data \cite{ATLAS:2012al} in Fig.~\ref{fig:atlaspower} and concluded that the wimpy showers give a better description of the data. Wimpy showers were therefore used for the MC event generation in this study.

\end{document}